\documentclass[journal]{IEEEtran}
\newtheorem{definition}{Definition}
\newtheorem{lemma}{Lemma}
\newtheorem{theorem}{Theorem}

\newtheorem{proposition}{Proposition}

\usepackage{color}
\usepackage{amsmath}
\usepackage{amssymb}
\usepackage{graphicx}
\usepackage{algorithm}
\usepackage{algpseudocode}
\algnewcommand{\IIf}[1]{\State\algorithmicif\ #1\ \algorithmicthen}
\algnewcommand{\EndIIf}{\unskip\ \algorithmicend\ \algorithmicif}
\ifCLASSINFOpdf
\else
\fi
\begin{document}
%
\title{On the Structure and Computation of Random Walk  Times in  Finite Graphs}
%
%
%

\author{Andrew Clark,~\IEEEmembership{Member,~IEEE,} Basel Alomair,~\IEEEmembership{Senior Member,~IEEE,}
               Linda Bushnell,~\IEEEmembership{Fellow,~IEEE}
        and Radha Poovendran,~\IEEEmembership{Fellow,~IEEE}
\thanks{A. Clark is with the Department of Electrical and Computer Engineering, Worcester Polytechnic Institute, Worcester, MA, 01609. {\tt aclark@wpi.edu}}%
\thanks{B. Alomair is with the Center for Cybersecurity, King Abdulaziz City for Science and Technology, Riyadh, Saudi Arabia. {\tt alomair@kacst.edu.sa}}
\thanks{L. Bushnell and R. Poovendran are with the Department of Electrical and Computer Engineering, University of Washington, Seattle, WA, 98195-2500. {\tt\{lb2, rp3\}@uw.edu}}
\thanks{This work was supported by NSF grant CNS-1544173 and ONR grants N00014-16-1-2710 and N00014-17-1-2946.}}

\maketitle

\begin{abstract}
We consider random walks in which the walk originates in one set of nodes and then continues until it reaches one or more nodes in a target set. The time required for the walk to reach the target set is of interest in understanding the convergence of Markov processes, as well as applications in control, machine learning, and social sciences.  In this paper, we investigate the computational structure of the random walk times as a function of the set of target nodes, and find that the commute, hitting, and cover times all exhibit submodular structure, even in non-stationary random walks. We provide a unifying proof of this structure by considering each of these times as special cases of stopping times. We generalize our framework to Markov Decision Processes, in which the target sets and control policies are jointly chosen to minimize the travel times, leading to polynomial-time approximation algorithms for choosing target sets. Our results are validated through numerical study.
\end{abstract}

\section{Introduction}
\label{sec:intro}
A random walk is a stochastic process over a graph, in which each node transitions to one of its neighbors at each time step according to a (possibly non-stationary) probability distribution \cite{levin2009markov,lovasz1993random}. Random walks on graphs are used to model and design  a variety of stochastic systems. Opinion propagation in social networks, as well as gossip and consensus algorithms in communications and networked control systems, are modeled and analyzed via random walks~\cite{ghaderi2013opinion,boyd2006randomized,jadbabaie2003coordination,sarkar2011random}. Random walks also serve as distance metrics in clustering, image segmentation, and other machine learning applications \cite{khoa2011large,yen2007graph,fouss2007random}. The behavior of physical processes, such as heat diffusion and electrical networks, can also be characterized through equivalent random walks~\cite{lawler2010random, doyle1984random}.

One aspect of random walks that has achieved significant research attention is the expected time for the walk to reach a given node or set of nodes~\cite{brightwell1990maximum,feige1995tight}. Relevant metrics include the hitting time, defined as the expected time for a random walk to reach any node in a given set, the commute time, defined as the expected time for the walk to reach any node in a given set and then return to the origin of the walk, and the cover time, which is the time for the walk to reach all nodes in a desired set. These times give rise to bounds on the rate at which the walk converges to a stationary distribution~\cite{levin2009markov}, and also provide metrics for quantifying the centrality or distance between nodes~\cite{fouss2007random,yen2008family}.  


The times of a random walk are related to system performance in a diverse set of network applications. The convergence rate of gossip and consensus algorithms is determined by the hitting time to a desired set of leader or anchor nodes~\cite{hunt2016algorithm,clark2014supermodular}. The effective resistance of an electrical network is captured by its commute time to the set of grounded nodes~\cite{chandra1996electrical}.  The performance of random-walk query processing algorithms is captured by the cover time of the set of nodes needed to answer the query~\cite{avin2004efficient}.  Optimal control problems such as motion planning and traffic analysis are described by the probability of reaching a target set or the resource cost per cycle of reaching a target set infinitely often~\cite{ding2011mdp}.



In each of these application domains, a set of nodes is selected in order to optimize one or more random walk metrics. The optimization problem will depend on whether the distribution of the random walk is affected by the choice of input nodes. Some systems will have a fixed walk distribution, determined by physical laws (such as heat diffusion or social influence propagation), for any set of input nodes. In other applications, the distribution can be selected by choosing a control policy, and hence the distribution of the walk and the set to be reached can be jointly optimized, as in Markov decision processes. In both cases, however, the number of possible sets of nodes is exponential in the network size, making the optimization problem computationally intractable in the worst case, requiring additional structure of the random walk times. At present, computational structures of random walks have received little attention by the research community. 


In this paper, we investigate the hitting, commute, cover, and cycle times, as well as the reachability probability, of a random walk as functions of the set of network nodes that are reached by the walk. We show that these metrics exhibit a submodular structure, and give a unifying proof of submodularity for the different metrics with respect to the chosen set of nodes. We consider both fixed random walk transition distributions and the case where the set of nodes and control policy  are jointly selected. We make the following specific contributions:
\begin{itemize}
\item We formulate the problem of jointly selecting a set of nodes and a control policy in order to maximize the probability of reaching the set from any initial state, as well as the average (per cycle) cost of reaching the set infinitely often. 

\item We prove that each problem can be approximated in polynomial time with provable optimality bounds using submodular optimization techniques. Our approach is to relate the existence of a probability distribution that satisfies a given cycle cost to the volume of a linear polytope, which is a submodular function of the desired set. We extend our approach to joint optimization of reachability and cycle cost, which we prove is equivalent to a matroid optimization problem.

\item In the case where the probability distribution is fixed, we develop a unifying framework, based on the submodularity of selecting subsets of stopping times, which includes proofs of the supermodularity of hitting and commute times and the submodularity of the cover time as special cases. Since the cover time is itself NP-hard to compute, we study and prove submodularity of standard upper bounds for the cover time.

\item We evaluate our results through numerical study, and show that the submodular structure of the system enables improvement over other heuristics such as greedy and centrality-based algorithms.
\end{itemize}

This paper is organized as follows. In Section \ref{sec:related}, we review the related work. In Section \ref{sec:preliminaries}, we present our system model and background on submodularity. In Section \ref{sec:optimal}, we demonstrate submodularity of random walk times when the walk distribution is chosen to optimize the times. In Section \ref{sec:fixed}, we study submodularity of random walk times with fixed distribution. In Section \ref{sec:simulation}, we present numerical results. In Section \ref{sec:conclusion}, we conclude the paper.

\section{Related Work}
\label{sec:related}
Commute, cover, and hitting times have been studied extensively, dating to classical bounds on the mixing time \cite{aldous2002reversible,levin2009markov}. Generalizations to these times have been proposed in \cite{coppersmith1993collisions,banderier2000generalized}. Connections between random walk times and electrical resistor networks were described in \cite{chandra1996electrical}. These classical works, however, do not consider the submodularity of the random walk times.

Submodularity of random walk times has been investigated based on connections between the hitting and commute times of random walks, and the performance of linear networked systems \cite{clark2014supermodular,clark2014minimizing}. The supermodularity of the commute time was shown in \cite{clark2014supermodular}. The supermodularity of the hitting time was shown in \cite{clark2014minimizing} and further studied in \cite{hunt2016algorithm}. These works assumed a fixed, stationary transition probability distribution, and also did not consider submodularity of the cover time. Our framework derives these existing results as special cases of a more general result, while also considering non-stationary transition probability distributions.  

Random walk times have been used for image segmentation and clustering applications. In these settings, the distance between two locations in an image is quantified via the commute time of a random walk with a probability distribution determined by a heat kernel \cite{yen2007graph}. Clustering algorithms were then proposed based on minimizing the commute time between two sets \cite{qiu2007clustering}.

This work is related to the problem of selecting an optimal control policy for a Markov decision process~\cite{fu2016optimal}. Prior work has investigated selecting a control policy in order to maximize the probability of reaching a desired target set \cite{papadimitriou1987complexity,maiga2016comprehensive}, or to minimize the cost per cycle of reaching the target set \cite{belta2016formal}. These works assume that the target set is given. In this paper, we consider the dual problem of selecting a target set in order to optimize these metrics.
\section{Background and Preliminaries}
\label{sec:preliminaries}
This section gives background on random walk times, Markov decision processes, and submodularity. Notations used throughout the paper are introduced.

\subsection{Random Walk Times}
\label{subsec:RW}

Throughout the paper, we let $\mathbf{E}(\cdot)$ denote the expectation of a random variable. Let $G = (V,E)$ denote a graph with vertex set $V$ and edge set $E$.  A random walk is a discrete-time random process $X_{k}$ with state space $V$. The distribution of the walk is defined by a set of functions $P_{k} : V^{k} \rightarrow \Pi^{V}$, where $$V^{k} = \underbrace{V \times \cdots \times V}_{k \mbox{ times}}$$ and $\Pi^{V}$ is the simplex of probability distributions over $V$. The function $P_{k}$ is defined by $$P_{k}(v_{1},\ldots,v_{k}) = Pr(X_{k} = v_{k} | X_{1} = v_{1}, \cdots, X_{k-1} = v_{k-1}).$$ The random walk is Markovian if, for each $k$, there exists a stochastic matrix $P_{k}$ such that $Pr(X_{k} = v_{k} | X_{1} = v_{1},\ldots,X_{k-1} = v_{k-1}) = P_{k}(v_{k-1},v_{k})$. Matrix $P_{k}$ is denoted as the transition matrix at time $k$. The random walk is Markovian and stationary if $P_{k} \equiv P$ for all $k$ and some stochastic matrix $P$. A Markovian and stationary walk is \emph{ergodic}~\cite{aldous2002reversible} if there exists a probability distribution $\pi$ such that, for any initial distribution $\phi$, $\lim_{k \rightarrow \infty}{\phi P^{k}} = \pi$, implying that the distribution of the walk will eventually converge to $\pi$ regardless of the initial distribution. 

Let $S \subseteq V$ be a subset of nodes in the graph. The hitting, commute, and cover time of $S$ are defined as follows. 

\begin{definition}
\label{def:times}
Let random variables $\nu(S)$, $\kappa(S)$, and $\phi(S)$ be defined as 
\begin{eqnarray*}
\nu(S) &=& \min{\{k : X_{k} \in S\}}\\
\kappa(S) &=& \min{\{k : X_{k} = X_{1}, X_{j} \in S \mbox{ for some } j < k\}} \\
\phi(S) &=& \min{\{k : \forall s \in S, \ \exists j \leq k \mbox{ s.t. } X_{j} = s\}}
\end{eqnarray*}
The hitting time of $S$ from a given node $v$ is equal to  $H(v,S) \triangleq \mathbf{E}(\nu(S) | X_{1} = v)$. The commute time of $S$ from node $v$ is equal to  $K(v,S) = \mathbf{E}(\kappa(S) | X_{1} = v)$. The cover time of $S$ from node $v$ is equal to  $C(v,S) = \mathbf{E}(\phi(S) | X_{1} = v)$. 
\end{definition}

Intuitively, the hitting time is the expected time for the random walk to reach the set $S$, the commute time is the expected time for a walk starting at a node $v$ to reach any node in $S$ and return to $v$, and the cover time is the expected time for the walk to reach all nodes in the set $S$. If $\pi$ is a probability distribution over $V$, we can further generalize the  above definitions to $H(\pi,S)$, $K(\pi,S)$, and $C(\pi,S)$, which are the expected hitting, commute, and cover times when the initial state is chosen from distribution $\pi$. 
The times described above are all special cases of stopping times of a stochastic process. 

\begin{definition}[\cite{levin2009markov}, Ch. 1]
\label{def:stopping-time}
A stopping time $Z$ of a random walk $X_{k}$ is a $\{1,2,\ldots,\infty\}$-valued random variable such that, for all $k$, the event $\{Z = k\}$ is determined by $X_{1},\ldots,X_{k}$. 
\end{definition}

 The hitting, cover, and commute times are stopping times.


\subsection{Markov Decision Processes}
\label{subsec:MDP}

A Markov Decision Process (MDP) is a generalization of a Markov chain, defined as follows.

\begin{definition}
\label{def:MDP}
An MDP $\mathcal{M}$ is a discrete-time random process $X_{k}$ defined by a tuple $\mathcal{M} = (V, \{A_{i} : i \in V\}, P)$, where $V$ is a set of states, $A_{i}$ is a set of actions at state $i$, and $P$ is a transition probability function defined by $$P(i,a,j) = Pr(X_{k+1} = j | X_{k} = i, \mbox{action $a$ chosen at time $k$}).$$
\end{definition}

Define $\mathcal{A} = \bigcup_{i=1}^{n}{A_{i}}$ and $A = |\mathcal{A}|$.  When the action set of a state is empty, the transition probability is a function of the current state only. 
The set of actions at each state corresponds to the possible control inputs that can be supplied to the MDP at that state. A \emph{control policy} is a function that takes as input a sequence of states $X_{1},\ldots,X_{k}$ and gives as output an action $a_{k} \in A_{X_{k}}$. A stationary control policy is a control policy $\mu$ that depends only on the current state, and hence can be characterized by a function $\mu: V \rightarrow A$. We let $\mathcal{P}$ denote the set of valid policies, i.e., the policies $\mu$ satisfying $\mu(i) \in A_{i}$ for all $i$. The random walk induced by a stationary policy $\mu$ is a stationary random walk with transition matrix $P_{\mu}$ defined by $P_{\mu}(i,j) = P(i,\mu(i),j)$.

The control policy $\mu$ is selected in order to achieve a design goal of the system. Such goals are quantified via specifications on the random process $X_{k}$. Two relevant specifications are \emph{safety} and \emph{liveness} constraints. A safety constraint specifies that a set of states $R$ should never be reached by the walk. A liveness constraint specifies that a given set of states $S$ must be reached infinitely often. In an MDP, two optimization problems arise in order to satisfy such constraints, namely, the reachability and average-cost-per-cycle problems.

The reachability problem consists of selecting a policy in order to maximize the probability that a desired set of states $S$ is reached by the Markov process while the unsafe set $R$ is not reached. The average cost per cycle (ACPC) problem is defined via the following metric. 
\begin{definition}
\label{def:ACPC}
The average cost per cycle metric from  initial state  $s \in V$ under policy $\mu$ is defined by 
\begin{equation}
\label{eq:ACPC}
J_{\mu}(s) = \limsup_{N \rightarrow \infty}{\mathbf{E}\left\{\frac{\sum_{k=0}^{N}{g(X_{k},\mu_{k}(X_{k}))}}{C(\mu,N)}\right\}},
\end{equation}

where $g(X_{k},\mu_{k}(X_{k}))$ is the cost of taking an action $\mu_{k}(X_{k})$ at state $X_{k}$ and $C$ is the number of cycles up to time $N$.
\end{definition}

The average cost per cycle can be viewed as the average number of steps in between times when the set $S$ is reached. The ACPC problem consists of choosing the set $S$ and policy $\mu$ in order to minimize $J(s)$. Applications of this problem include motion planning, in which the goal is to reach a desired state infinitely often while minimizing energy consumption.

We define  $J_{\mu} \in \mathbb{R}^{n}$ as the vector of ACPC values for different initial states, so that $J_{\mu}(s)$ is the ACPC value for state $s$. In the special case where all actions have cost $1$, Eq. (\ref{eq:ACPC}) is equivalent to
\begin{equation}
\label{eq:ACPC-same-cost}
J_{\mu}(s) = \limsup_{N \rightarrow \infty}{\left\{\frac{N}{C(\mu,N)}\right\}}.
\end{equation}
We focus on this case in what follows. It has been shown in \cite{ding2014optimal} that the optimal policy $\mu^{\ast}$ that minimizes the ACPC is independent of the initial state. The following theorem characterizes the minimum ACPC policy $\mu^{\ast}$.

\begin{theorem}[\cite{ding2014optimal}]
\label{theorem:min-ACPC}
The optimal ACPC is given by $J_{\mu^{\ast}} = \lambda\mathbf{1}$, where $\lambda \in \mathbb{R}$ and there exist vectors $h$ and $\nu$ satisfying 
\begin{eqnarray}
J_{\mu^{\ast}} + h &=& \mathbf{1} + P_{\mu^{\ast}}h + \overline{P}_{\mu^{\ast}}J_{\mu^{\ast}} \\
P_{\mu^{\ast}}   &=& (I-\overline{P}_{\mu^{\ast}})h + \nu  \\
\nonumber
\lambda + h(i) &=& \min_{a \in A_{i}}{\left[1 + \sum_{j=1}^{n}{P(i,a,j)h(j)}\right.} \\
\label{eq:ACPC-optimal}
&& \left. \qquad + \lambda \sum_{j \notin S}{P(i,a,j)}\right].
\end{eqnarray}
$P_{\mu^{\ast}}$ is the transition matrix induced by $\mu^{\ast}$ and 
\begin{equation}
\overline{P}_{\mu^{\ast}}(i,j) = \left\{
\begin{array}{ll}
P_{\mu}(i,j), & j \notin S \\
0, \mbox{else}
\end{array}
\right.
\end{equation}
\end{theorem}

\subsection{Submodularity}
\label{subsec:submod}
Submodularity is a property of set functions $f: 2^{W} \rightarrow \mathbb{R}$, where $W$ is a finite set and $2^{W}$ is the set of all subsets of $W$. A function is submodular if, for any sets $S$ and $T$ with $S \subseteq T \subseteq W$ and any $v \notin T$, $$f(S \cup \{v\}) - f(S) \geq f(T \cup \{v\}) - f(T).$$ A function is supermodular if $-f$ is submodular, while a function is modular if it is both submodular and supermodular. For any modular function $f(S)$, a set of coefficients $\{c_{i} : i \in W\}$ can be defined such that $$f(S) = \sum_{i \in S}{c_{i}}.$$ Furthermore, for any set of coefficients $\{c_{i} : i \in W\}$, the function $f(S) = \max{\{c_{i} : i \in S\}}$ is increasing and submodular, while the function $\min{\{c_{i} : i \in S\}}$ is decreasing and supermodular. Any nonnegative weighted sum of submodular (resp. supermodular) functions is submodular (resp. supermodular).

A matroid is defined as follows. 

\begin{definition}
\label{def:matroid}
A matroid $\mathcal{M}=(V,\mathcal{I})$ is defined by a set $V$ and a collection of subsets $\mathcal{I}$. The set $\mathcal{I}$ satisfies the following conditions: (i) $\emptyset \in \mathcal{I}$, (ii) $B \in \mathcal{I}$ and $A \subseteq B$ implies that $A \in \mathcal{I}$, and (iii) If $|A| < |B|$ and $A, B \in \mathcal{I}$, then there exists $v \in B \setminus A$ such that $(A \cup \{v\}) \in \mathcal{I}$.
\end{definition}

The collection of sets $\mathcal{I}$ is denoted as the independent sets of the matroid. A \emph{basis} is a maximal independent set. The uniform matroid $\mathcal{M}_{k}$ is defined by $A \in \mathcal{I}$ iff $|A| \leq k$. A \emph{partition matroid} is defined as follows.

\begin{definition}
\label{def:partition}
Let $V = V_{1} \cup \cdots \cup V_{m}$ with $V_{i} \cap V_{j} = \emptyset$ for $i \neq j$ be a partition of a set $V$. The partition matroid $\mathcal{M} = (V, \mathcal{I})$ is defined by $A \in \mathcal{I}$ if $|A \cap V_{i}| \leq 1$ for all $i=1,\ldots,m$.
\end{definition}

Finally, given two matroids $\mathcal{M}_{1} = (V, \mathcal{I}_{1})$ and $\mathcal{M}_{2} = (V, \mathcal{I}_{2})$, the union $\mathcal{M} = \mathcal{M}_{1} \vee \mathcal{M}_{2}$ is a matroid in which $A \in \mathcal{I}$ iff $A = A_{1} \cup A_{2}$ for some $A_{1} \in \mathcal{I}_{1}$ and $A_{2} \in \mathcal{I}_{2}$.


\section{Random Walks on Markov Decision Processes}
\label{sec:optimal}

In this section, we consider the problem of selecting a set $S$ of states for an MDP to reach in order to optimize a performance metric. We consider two problems, namely, the problem of selecting a set of states $S$ in order to maximize the reachability probability to $S$ while minimizing the probability of reaching an unsafe set $R$, and the problem of selecting a set of states $S$ in order to minimize the ACPC. A motivating scenario is a setting where an unmanned vehicle must reach a refueling station or transit depot infinitely often, and the goal is to place the set of stations in order to maximize the probability of reaching one or minimize the cost of reaching.  

\subsection{Reachability Problem Formulation}
\label{subsec:reachability-formulation}


The problem of selecting a set $S$ with at most $k$ nodes in order to maximize the probability of reaching $S$ under the optimal policy $\mu$ is considered as follows. Let $\sigma(S)$ denote the event that the walk reaches $S$ at some finite time and does not reach the unsafe state $R$ at any time.  The problem formulation is given by 
\begin{equation}
\label{eq:reachability-formulation}
\begin{array}{ll} 
\mbox{maximize} & \max_{\mu \in \mathcal{P}}{Pr(\sigma(S)|\mu)} \\
S \subseteq V & \\
\mbox{s.t.} & |S| \leq k
\end{array}
\end{equation}

Here $Pr(\sigma(S) | \mu)$ denotes the probability that $\sigma(S)$ occurs when the policy is $\mu$. This formulation is equivalent to 
\begin{equation}
\label{eq:equiv-reachability-formulation}
\begin{array}{lll} 
\mbox{maximize} &  \mbox{max} & Pr(\sigma(S)|\mu) \\
\mu & S: |S| \leq k & 
\end{array}
\end{equation}

The following known result gives a linear programming approach to maximizing the probability of reachability for a fixed set $S$.

\begin{lemma}[\cite{baier2008principles}, Theorem 10.105]
\label{lemma:fixed-S-reachability}
The optimal value of $\max{\{Pr(\sigma(S)|\mu) : \mu \in \mathcal{P}\}}$ is equal to 

\begin{equation}
\label{eq:fixed-S-reachability}
\begin{array}{ll} 
\mbox{min} & \mathbf{1}^{T}\mathbf{x} \\
\mathbf{x} \in \mathbb{R}^{n} & \\
\mbox{s.t.} & x_{i} \in [0,1] \ \forall i \\
 &  x_{i} = 1 \ \forall i \in S, x_{i} = 0 \ \forall i \in R \\
 & x_{i} \geq \sum_{j=1}^{n}{P(i,a,j)x_{j}} \ \forall i \in V \setminus R, a \in A_{i} 
\end{array}
\end{equation}
\end{lemma}


The optimal solution $\mathbf{x}$ to the linear program (\ref{eq:fixed-S-reachability}) is a vector in $\mathbb{R}^{n}$, where $x_{i}$ is the probability that $\sigma(S)$ occurs under the optimal policy when the initial state is $i$. 

In addition to giving the optimal value of the maximal reachability problem, Eq. (\ref{eq:fixed-S-reachability}) can also be used to compute the optimal policy. In order for $\mathbf{x}$ to be the solution to (\ref{eq:fixed-S-reachability}), for each $i \in V \setminus (R \cup S)$, there must be an action $a_{i}^{\ast}$ such that $$x_{i} = \sum_{j=1}^{n}{P(i,a_{i}^{\ast},j)x_{j}}.$$ Otherwise, it would be possible to decrease $x_{i}$ and hence the objective function of (\ref{eq:fixed-S-reachability}). Hence the optimal policy $\mu$ is given by $\mu(i) = a_{i}^{\ast}$. 

We first define a relaxation of (\ref{eq:fixed-S-reachability}). Let $\rho > 0$, and define the relaxation by
\begin{equation}
\label{eq:relaxed-fixed-opt}
\begin{array}{ll}
\mbox{minimize} & \mathbf{1}^{T}\mathbf{x} + \rho\left(\sum_{i \in S}{(1-x_{i})}\right) \\
\mathbf{x} & \\
\mbox{s.t.} & \mathbf{x} \in [0,1]^{n} \\
 & x_{i} = 0 \ \forall i \in R \\
 & x_{i} \geq \sum_{j=1}^{n}{P(i,a,j)x_{j}} \ \forall i \in V \setminus R, a \in A_{i}
\end{array}
\end{equation}

The following lemma shows that (\ref{eq:relaxed-fixed-opt}) is equivalent to (\ref{eq:fixed-S-reachability}).

\begin{lemma}
\label{lemma:relaxed-reachability-equivalent}
When $\rho > n$, the optimal solutions and optimal values of (\ref{eq:fixed-S-reachability}) and (\ref{eq:relaxed-fixed-opt}) are equal.
\end{lemma}

\begin{IEEEproof}
We first show that the solution to (\ref{eq:relaxed-fixed-opt}) satisfies $x_{i}=1$ for all $i \in S$. Suppose that this is not the case. Let $\mathbf{x}^{\ast}$ denote the solution to (\ref{eq:relaxed-fixed-opt}), and suppose that $x_{r}^{\ast} = 1-\epsilon$ for some $\epsilon > 0$ and $r \in S$. Now, construct a new vector $\mathbf{x}^{\prime} \in \mathbb{R}^{n}$ as
\begin{displaymath}
x_{i}^{\prime} = \left\{
\begin{array}{ll}
0, & i \in R \\
\min{\{x_{i}^{\ast} + \epsilon, 1\}}, & i \notin R
\end{array}
\right.
\end{displaymath}
Note that for all $i$, $0 \leq (x_{i}^{\prime} - x_{i}^{\ast}) \leq \epsilon$. We will first show that $\mathbf{x}^{\prime}$ is feasible under the constraints of (\ref{eq:relaxed-fixed-opt}), and then show that the resulting objective function value is less than the value produced by $\mathbf{x}^{\ast}$, contradicting optimality of $\mathbf{x}^{\ast}$. 

By construction, $\mathbf{x}^{\prime} \in [0,1]^{n}$ and $x_{i}^{\prime} = 0$ for all $i \in R$. For each $i \notin R$, suppose first that $x_{i}^{\prime} = 1$. Then $$x_{i}^{\prime} = 1 = \sum_{j=1}^{n}{P(i,a,j)1} \geq \sum_{j=1}^{n}{P(i,a,j)x_{j}^{\prime}}.$$ Suppose next that $x_{i}^{\prime} = x_{i} + \epsilon$. Then for all $a \in A_{i}$,
\begin{IEEEeqnarray*}{rCl}
\sum_{j=1}^{n}{P(i,a,j)x_{j}^{\prime}} &=& \sum_{j=1}^{n}{P(i,a,j)(x_{j}^{\ast} + (x_{j}^{\prime}-x_{j}^{\ast}))} \\
&=& \sum_{j=1}^{n}{P(i,a,j)x_{j}^{\ast}} + \sum_{j=1}^{n}{P(i,a,j)(x_{j}^{\prime}-x_{j}^{\ast})} \\
&\leq& \sum_{j=1}^{n}{P(i,a,j)x_{j}^{\ast}} + \epsilon \sum_{j=1}^{n}{P(i,a,j)} \\
&\leq& x_{i}^{\ast} + \epsilon = x_{i}^{\prime}
\end{IEEEeqnarray*}
implying that $\mathbf{x}^{\prime}$ is feasible. The objective function value of $\mathbf{x}^{\prime}$ is given by 
\begin{IEEEeqnarray*}{rCl}
\mathbf{1}^{T}\mathbf{x}^{\prime} + \rho \left(\sum_{i \in S}{(1-x_{i}^{\prime})}\right) &=& \mathbf{1}^{T}\mathbf{x}^{\ast} + \mathbf{1}^{T}(\mathbf{x}^{\prime}-\mathbf{x}^{\ast}) \\
&& + \rho\left(\sum_{i \in S \setminus \{r\}}{(1-x_{i}^{\prime})}\right) \\
&\leq& \mathbf{1}^{T}\mathbf{x}^{\ast} + \epsilon n + \rho\left(\sum_{i \in S \setminus \{r\}}{(1-x_{i}^{\ast})}\right) \\
&<& \mathbf{1}^{T}\mathbf{x}^{\ast} + \epsilon\rho + \rho\sum_{i \in S \setminus \{r\}}{(1-x_{i}^{\ast})} \\
&=& \mathbf{1}^{T}\mathbf{x}^{\ast} + \rho\sum_{i \in S}{(1-x_{i}^{\ast})}
\end{IEEEeqnarray*}
contradicting the assumption that $\mathbf{x}^{\ast}$ is the optimal value of (\ref{eq:relaxed-fixed-opt}). Hence the optimal value of (\ref{eq:relaxed-fixed-opt}) minimizes $\mathbf{1}^{T}\mathbf{x}$ while satisfying $x_{i} = 1$ for all $i \in S$, which is equivalent to the solution of (\ref{eq:fixed-S-reachability}).
\end{IEEEproof}

The problem of maximizing reachability is therefore equivalent to 

\begin{equation}
\label{eq:relaxed-opt}
\begin{array}{lll}
\mbox{maximize} & \mbox{min} & \mathbf{1}^{T}\mathbf{x} + \rho \left(\sum_{i \in S}{(1-x_{i})}\right) \\
S: |S| \leq k & \mathbf{x} \in \Pi & 
\end{array}
\end{equation}

The min-max inequality implies that (\ref{eq:relaxed-opt}) can be bounded above by 

\begin{equation}
\label{eq:reachability-tractable}
\begin{array}{lll}
\mbox{min} & \mbox{max} & \mathbf{1}^{T}\mathbf{x} + \rho \left(\sum_{i \in S}{(1-x_{i})}\right) \\
\mathbf{x} \in \Pi & S: |S| \leq k &   
\end{array}
\end{equation}

The objective function of (\ref{eq:reachability-tractable}) is a pointwise maximum of convex functions and is therefore convex. A subgradient of the objective function at any point $\mathbf{x}_{0}$, denoted $v(\mathbf{x}_{0})$, is given by 
\begin{displaymath}
v(\mathbf{x}_{0})_{i} = \left\{
\begin{array}{ll}
1-\rho, & i \in S_{max}(\mathbf{x}_{0}) \\
1, & \mbox{else}
\end{array}
\right.
\end{displaymath}
where $$S_{max}(\mathbf{x}_{0}) = \arg\min{\left\{\sum_{i \in S}{(x_{0})_{i}} : |S| \leq k\right\}}.$$ This subgradient can be computed efficiently by selecting the $k$ largest elements of $\mathbf{x}_{0}$.  A polynomial-time algorithm for solving (\ref{eq:reachability-tractable}) can be obtained using interior-point methods, as shown in Algorithm \ref{algo:reachability}.

   \begin{center}
\begin{algorithm}[!htp]
	\caption{Algorithm for selecting a set of states $S$ to maximize probability of reachability.}
	\label{algo:reachability}
	\begin{algorithmic}[1]
		\Procedure{Max\_Reach}{$G=(V,E)$, $A$, $P$, $R$, $k$, $\epsilon$, $\delta$}
            \State \textbf{Input}: Graph $G=(V,E)$, set of actions $A_{i}$ at each state $i$, probability distribution $P$, unsafe states $R$, number of states $k$, convergence parameters $\epsilon$ and $\delta$.
            \State \textbf{Output}: Set of nodes $S$
            \State $\Phi \leftarrow$ barrier function for polytope $\Pi$
            \State $\mathbf{x} \leftarrow 0$
            \State $\mathbf{x}^{\prime} \leftarrow \mathbf{1}$
            \While{$||\mathbf{x}-\mathbf{x}^{\prime}||_{2} > \epsilon$}
            \State $S \leftarrow \arg\max{\{\sum_{i \in S}{x_{i}}: |S| \leq k\}}$
            \State $v \leftarrow 1$
            \State $v_{i} \leftarrow (1-\rho) \ \forall i \in S$
            \State $w \leftarrow \nabla_{\mathbf{x}}{\Phi(\mathbf{x})}$
            \State $\mathbf{x}^{\prime} \leftarrow \mathbf{x}$
            \State $\mathbf{x} \leftarrow \mathbf{x} + \delta (v + w)$
            \EndWhile
            \State $S \leftarrow \arg\max{\{\sum_{i \in S}{x_{i}}: |S| \leq k\}}$
            \State \Return{$S$}
		\EndProcedure
	 \end{algorithmic}
\end{algorithm}
\end{center}


The interior-point approach of Algorithm \ref{algo:reachability} gives an efficient algorithm for maximizing the probability of reaching $S$. We further observe that more general constraints than $|S| \leq k$ can be constructed. One possible constraint is to ensure that, for some partition $V_{1}, \ldots, V_{m}$, we have $|S \cap V_{i}| \geq 1$ for each $i=1,\ldots,m$. Intuitively, this constraint implies that there must be at least one state to be reached in each partition set $V_{i}$. This constraint is equivalent to the constraint $S \in \mathcal{M}_{k}$, where $\mathcal{M}_{k}$ is the union of the partition matroid and the uniform matroid of rank $k-m$. The calculation of $S_{max}(\mathbf{x}_{0})$ then becomes $$S_{max}(\mathbf{x}_{0}) = \arg\min{\left\{\sum_{i \in S}{(x_{0})_{i}} : S \in \mathcal{M}_{k}\right\}}.$$ This problem can also be solved in polynomial time due to the matroid structure of $\mathcal{M}_{k}$ using a greedy algorithm.


\subsection{Minimizing the Average Cost Per Cycle}
\label{subsec:ACPC}
This section considers the problem of selecting a set $S$ in order to minimize the ACPC. 
%
Based on Theorem \ref{theorem:min-ACPC}, in order to ensure that the minimum ACPC is no more than $\lambda$, it suffices to show that there is no $h$ satisfying (\ref{eq:ACPC-optimal}) for the chosen set $S$. Note that this condition is sufficient but not necessary. Hence the following optimization problem gives a lower bound on the ACPC 

\begin{equation}
\label{eq:ACPC-opt}
\begin{array}{lll}
\mbox{minimize} & \mbox{max} & \lambda \\
S & \lambda, h & \\
 & 
\mbox{s.t.} & \lambda + h(i) \leq 1 + \sum_{j=1}^{n}{P(i,a,j)h(j)} \\ 
& & +\lambda\sum_{j \notin S}{P(i,a,j)} \ \forall i, a \in A_{i} \\
\end{array}
\end{equation}

The following theorem gives a sufficient condition for the minimum ACPC.

\begin{theorem}
\label{theorem:ACPC-sufficient}
Suppose that, for any $h \in \mathbb{R}^{n}$, there exist $a$ and $i$ such that
\begin{equation}
\label{eq:ACPC-sufficient}
\lambda\mathbf{1}\{i \in S\} + h(i) - \sum_{j=1}^{n}{P(i,a,j)h(j)} > 1.
\end{equation}
Then the ACPC is bounded above by $\lambda$.
\end{theorem}

 In order to prove Theorem \ref{theorem:ACPC-sufficient}, we introduce an extended state space that will have the same ACPC. The state space is defined $\hat{V} = V \times \{0,1\}$.  The sets of actions satisfy $\hat{A}_{(i,0)} = \hat{A}_{(i,1)} = A_{i}$. The transition probabilities are given by $$\hat{P}((i,1),a,(i,0)) = 1, \quad \hat{P}((i,0),a,(j,1)) = P(i,j).$$ Finally, the set $\hat{S}$  is constructed from the set $S$  via $$\hat{S} = \{(i,0) : i \in S\}.$$  The following result establishes the equivalence between the graph formulations.

\begin{proposition}
\label{prop:ACPC-equivalent}
The minimum ACPC of $\mathcal{M}$ with set $S$ is equal to the minimum ACPC of $\hat{\mathcal{M}} = (\hat{V}, \hat{A}, \hat{P})$ with set $\hat{S}$.
\end{proposition}

\begin{IEEEproof}
There is a one-to-one correspondence between policies on $\mathcal{M}$ and policies on $\hat{\mathcal{M}}$. Indeed, any policy $\mu$ on $\mathcal{M}$ can be extended to a policy $\hat{\mu}$ on $\hat{\mathcal{M}}$ by setting $\hat{\mu}(i,0) = \mu(i)$  and $\hat{\mu}(i,1) = 1$ for all $i$.  Furthermore, by construction, the ACPC for $\mathcal{M}$ with policy $\mu$ will be equal to the ACPC with policy $\hat{\mu}$. In particular, the cost per cycle of the minimum-cost policies will be equal.
\end{IEEEproof}

We are now ready to prove Theorem \ref{theorem:ACPC-sufficient}.

\begin{IEEEproof}[Proof of Theorem \ref{theorem:ACPC-sufficient}]
For the MDP $\hat{\mathcal{M}}$, the constraint of Eq. (\ref{eq:ACPC-opt}) is equivalent to 
\begin{eqnarray}
\label{eq:ACPC-1}
\lambda \mathbf{1}\{i \in S\} + \hat{h}(i,1) &\leq& \hat{h}(i,0) \\
\label{eq:ACPC-2}
\hat{h}(i,0) - \sum_{j=1}^{n}{P(i,a,j)\hat{h}(j,1)} &\leq& 1
\end{eqnarray}
for all $i$, $j$, and $u$. Hence, in order for the minimum cost per cycle to be less than $\lambda$, at least one of (\ref{eq:ACPC-1}) or (\ref{eq:ACPC-2}) must fail for each $\hat{h} \in \mathbb{R}^{2N}$. For each $\hat{h}$, let  $S_{\hat{h}} = \{i: \lambda + \hat{h}(i,1) > \hat{h}(i,0)\}$, so that the condition that the ACPC is less than $\lambda$ is equivalent to $$S_{\hat{h}} \cap S \neq \emptyset \quad \forall \hat{h} \in \mathbb{R}^{2N}.$$ Furthermore, we can combine Eq. (\ref{eq:ACPC-1}) and (\ref{eq:ACPC-2}) to obtain  $$\lambda\mathbf{1}\{i \in S\} + \hat{h}(i,1) \leq 1 + \sum_{j=1}^{n}{P(i,a,j)\hat{h}(j,1)}.$$ For the MDP $\mathcal{M}$, define $h \in \mathbb{R}^{N}$ by $h(i) = h(i,1)$ for all $i \in V$, so that  
\begin{equation}
\label{eq:another-ACPC}
\lambda\mathbf{1}\{i \in S\} + h(i) \leq 1 + \sum_{j=1}^{n}{P(i,a,j)h(j)}.
\end{equation}
Since $\mathcal{M}$ and $\hat{\mathcal{M}}$ have the same ACPC, (\ref{eq:another-ACPC}) is a necessary condition for the ACPC of $\mathcal{M}$ to  be at least $\lambda$. This is equivalent to (\ref{eq:ACPC-sufficient}).
\end{IEEEproof}
 We will now map the minimum ACPC problem to submodular optimization. As a preliminary, define the matrix $\overline{A} \in \mathbb{R}^{nA \times n}$, with rows indexed $\{(i,a) : i \in 1,\ldots,n, a \in \mathcal{A}\}$,  as 
 \begin{displaymath}
 \overline{A}((i,a),j) = \left\{
 \begin{array}{ll}
 -P(i,a,j), & i \neq j \\
 1-P(i,a,j), & i = j
 \end{array}
\right.
 \end{displaymath}
and the vector $b(S) \in \mathbb{R}^{nA}$, with entries indexed $\{(i,a): i =1,\ldots,n, a \in \mathcal{A}\}$, as 
\begin{displaymath}
(b(S))_{i,a} = \left\{
\begin{array}{ll}
1-\lambda, & i \in S \\
1, & i \notin S
\end{array}
\right.
\end{displaymath}
 

\begin{proposition}
\label{prop:ACPC-submod}
For any $\zeta > 0$, let $\mathcal{P}(\lambda,S)$ denote the polytope $$\mathcal{P}(\lambda,S) = \{h: \overline{A}h \leq b(S)\} \cap \{||h||_{\infty} \leq \zeta\}.$$ Then the function $r_{\lambda}(S) = \mbox{vol}(\mathcal{P}(\lambda,S))$ is decreasing and supermodular as a function of $S$. Furthermore, if $r_{\lambda}(S) = 0$, then the ACPC is bounded above by $\lambda$.
\end{proposition}

\begin{IEEEproof}
Define a sequence of functions $r^{N}_{\lambda}(S)$ as follows. For each $N$, partition the set $\mathcal{P}(0, \emptyset) = \{h : \overline{A}h \leq \mathbf{1}\}$ into $N$ equally-sized regions with center $x_{1},\ldots,x_{N}$ and  volume $\delta_{N}$. Define $$r^{N}_{\lambda}(S) = \sum_{l=1}^{N}{\delta_{N}\mathbf{1}\{x_{l} \in \mathcal{P}(\lambda, S)\}}.$$ 
Since $\mathcal{P}(\lambda,S) \subseteq \mathcal{P}(0,\emptyset)$ for all $S$ and $\lambda$, we have that $$\mbox{vol}(\mathcal{P}(\lambda,S)) \approx r^{N}_{\lambda}(S)$$ and $$\lim_{N \rightarrow \infty}{r^{N}_{\lambda}(S)} = r_{\lambda}(S).$$ 
The term $\mathbf{1}\{x_{l} \in \mathcal{P}(\lambda,S)\}$ is equal to the decreasing supermodular function $$\mathbf{1}\{S_{x_{l}} \cap S = \emptyset \}.$$ Hence $r^{N}_{\lambda}(S)$ is a decreasing supermodular function, and $r_{\lambda}(S)$ is a limit of decreasing supermodular functions and is therefore decreasing supermodular. Finally, we have that if $r_{\lambda}(S) = 0$, then  there is no $h$ satisfying $\overline{A}h \leq b(S)\}$, and hence the ACPC is bounded above by $\lambda$ by Theorem \ref{theorem:ACPC-sufficient}.
\end{IEEEproof}

In Proposition \ref{prop:ACPC-submod}, the constraint $||h||_{\infty} \leq \zeta$ is added to ensure that the polytope is compact.
 
 Proposition \ref{prop:ACPC-submod} implies that ensuring that the ACPC is bounded above by $\lambda$ is equivalent to the submodular constraint $r_{\lambda}(S) = 0$.  Hence, $r_{\lambda}(S)$ is a submodular metric that can be used to ensure a given bound $\lambda$ on ACPC. This motivates a bijection-based algorithm for solving the minimum ACPC problem (Algorithm \ref{algo:ACPC}). 
 
    \begin{center}
\begin{algorithm}[!htp]
	\caption{Algorithm for selecting a set of states $S$ to minimize average cost per cycle.}
	\label{algo:ACPC}
	\begin{algorithmic}[1]
		\Procedure{Min\_ACPC}{$V$, $A$, $P$, $R$, $k$}
            \State \textbf{Input}:Set of states $V$, set of actions $A_{i}$ at each state $i$, probability distribution $P$, unsafe states $R$, number of states to be chosen $k$.
            \State \textbf{Output}: Set of nodes $S$
            \State $\lambda_{max} \leftarrow $ max ACPC for any $v \in \{1,\ldots,n\}$
            \State $\lambda_{min} \leftarrow 0$
            \While{$|\lambda_{max}-\lambda_{min}| > \delta$}
            \State $S \leftarrow \emptyset$
            \State $\lambda \leftarrow \frac{\lambda_{max}+\lambda_{min}}{2}$
            \While{$r_{\lambda}(S) > 0$}
            \State $v^{\ast} \leftarrow \arg\min{\{r_{\lambda}(S \cup \{v\}) : v \in \{1,\ldots,n\}\}}$
            \State $S \leftarrow S \cup \{v^{\ast}\}$
            \EndWhile
            \If{$|S| \leq k$}
            \State $\lambda_{max} \leftarrow \lambda$
            \Else
            \State $\lambda_{min} \leftarrow \lambda$
            \EndIf
            \EndWhile
            \State \Return{$S$}
		\EndProcedure
	 \end{algorithmic}
\end{algorithm}
\end{center}

The following theorem describes the optimality bounds and complexity of Algorithm \ref{algo:ACPC}.

\begin{theorem}
\label{theorem:ACPC-complexity}
Algorithm \ref{algo:ACPC} terminates in $O\left(kn^{6}\log{\lambda_{max}}\right)$ time. For any $\lambda$ such that there exists a set $S$ of size $k$ satisfying $r_{\lambda}(S) = 0$, Algorithm \ref{algo:ACPC} returns a set $S^{\prime}$ with $$\frac{|S^{\prime}|}{|S|} \leq 1 + \log{\left\{\frac{r_{\lambda}(\emptyset)}{\min_{v}{\{r_{\lambda}(\hat{S} \setminus \{v\})\}}}\right\}}.$$ 
\end{theorem}

\begin{IEEEproof}
The number of rounds in the outer loop is bounded by $\log{\lambda_{max}}$. For each iteration of the inner loop, the objective function $r_{\lambda}(S)$ is evaluated $kn$ times. Computing $r_{\lambda}(S)$ is equivalent to computing the volume of a linear polytope, which can be approximated in $O(n^{5})$ time~\cite{lovasz1993volume}, for a total runtime of $O(kn^{6}\log{\lambda_{max}})$. 

From \cite{wolsey1982analysis}, for any monotone submodular function $f(S)$ and the optimization problem $\min{\{|S| : f(S) \leq \alpha\}}$, the set $\hat{S}$ returned by the algorithm satisfies $$\frac{|\hat{S}|}{|S^{\ast}|} \leq 1 + \log{\left\{\frac{f(V) - f(\emptyset)}{f(V) - f(\hat{S}_{T-1})}\right\}},$$ where $\hat{S}_{T-1}$ is the set obtained at the second-to-last iteration of the algorithm. Applied to this setting, we have 
\begin{eqnarray*}
\frac{|\hat{S}|}{|S^{\ast}|} &\leq& 1 + \log{\left\{\frac{r_{\lambda}(V) - r_{\lambda}(\emptyset)}{r_{\lambda}(V) - r_{\lambda}(\hat{S}_{T-1})}\right\}} \\
&=& 1 + \log{\left\{\frac{r_{\lambda}(\emptyset)}{r_{\lambda}(\hat{S}_{T-1})}\right\}} \\
&\leq& 1 + \log{\left\{\frac{r_{\lambda}(\emptyset)}{\min_{v}{\{r_{\lambda}(\hat{S} \setminus \{v\})\}}}\right\}}.
\end{eqnarray*}
\end{IEEEproof}

We note that the complexity of Algorithm \ref{algo:ACPC} is mainly determined by the complexity of computing the volume of the polytope $\mathcal{P}(\lambda,S)$. This complexity can be reduced to $O(n^{3})$  by computing the volume of the minimum enclosing ellipsoid of $\mathcal{P}(\lambda,S)$ instead. 

The approach for minimizing ACPC presented in this section is applicable to problems such as motion planning for mobile robots. The set $S$ represents locations that must be reached infinitely often, as in a surveillance problem, while minimizing ACPC can be viewed as minimizing the total resource consumption (e.g., fuel costs) while reaching the desired states infinitely often. 


\subsection{Joint Optimization of Reachability and ACPC}
\label{subsec:joint-reach-ACPC}
In this section, we consider the problem of selecting a set $S$ to satisfy safety and liveness constraints with maximum probability and minimum cost per cycle. This problem can be viewed as combining the maximum reachability and minimum ACPC problems formulated in the previous sections. As a preliminary, we define an end component of an MDP.

\begin{definition}
\label{def:EC}
 An end component (EC) of an MDP $\mathcal{M} = (V, \mathcal{A}, P)$ is an MDP $\mathcal{M}^{\prime} = (V^{\prime}, \mathcal{A}^{\prime}, P^{\prime})$ where (i) $V^{\prime} \subseteq V$, (ii) $A_{i}^{\prime} \subseteq A_{i}$ for all $i \in V^{\prime}$, (iii) $P^{\prime}(i,a,j) = P(i,a,j)$ for all $i,j \in V^{\prime}$ and $a \in A_{i}^{\prime}$, and (iv) if $i \in V^{\prime}$ and $P(i,a,j) > 0$ for some $a \in A_{i}^{\prime}$, then $j \in V^{\prime}$.
 \end{definition}
 
 Intuitively, an end component $(V^{\prime}, \mathcal{A}^{\prime}, P^{\prime})$ is a set of states and actions such that if only the actions in $\mathcal{A}^{\prime}$ are selected, the MDP will remain in $V^{\prime}$ for all time. A maximal end component (MEC) $(V^{\prime}, \mathcal{A}^{\prime},P^{\prime})$ such that for any $V^{\prime\prime}$ and $\mathcal{A}^{\prime\prime}$ with $V^{\prime} \subseteq V^{\prime\prime}$ and $\mathcal{A}^{\prime} \subseteq \mathcal{A}^{\prime\prime}$, there is no EC with vertex set $V^{\prime\prime}$ and set of actions $\mathcal{A}^{\prime\prime}$.
 
 
 \begin{lemma}[\cite{ding2014optimal}]
\label{lemma:EC-safety-liveness}
The probability that an MDP satisfies a safety and liveness specification defined by $R$ and $S$ is equal to the probability that the MDP reaches an MEC $(V^{\prime}, \mathcal{A}^{\prime}, P^{\prime})$ with $V^{\prime} \cap S \neq \emptyset$ and $V^{\prime} \cap R = \emptyset$.
\end{lemma}

We define an MEC satisfying $V^{\prime} \cap R = \emptyset$ to be an accepting maximal end component (AMEC). By Lemma \ref{lemma:EC-safety-liveness}, the problem of maximizing the probability of satisfying the safety and liveness constraints is equal to the probability of reaching an AMEC with $V^{\prime} \cap S \neq \emptyset$. Let $\mathcal{M}_{1} = (V_{1}^{\prime}, A_{1}^{\prime}, P_{1}^{\prime}), \ldots, \mathcal{M}_{N}=(V_{N}^{\prime},A_{N}^{\prime},P_{N}^{\prime})$ denote the set of AMECs of $\mathcal{M}$ satisfying $V_{i}^{\prime} \cap S \neq \emptyset$.


We now formulate two problems of joint reachability and ACPC. The first problem is to minimize the ACPC, subject to the constraint that the probability of satisfying the constraints is maximized. The second problem is to maximize the probability of satisfying safety and liveness properties, subject to a constraint on the average cost per cycle. In order to address the first problem, we characterize the sets $S$ that maximize the reachability probability.

\begin{lemma}
\label{lemma:max_reach}
Suppose that for each AMEC $(V^{\prime}, \mathcal{A}^{\prime}, P^{\prime})$, $S \cap V^{\prime} \neq \emptyset$. Then $S$ maximizes the probability of satisfying the safety and liveness constraints of the MDP.
\end{lemma}

\begin{IEEEproof}
By Lemma \ref{lemma:EC-safety-liveness}, the safety and liveness constraints are satisfied if the walk reaches an MEC satisfying $S \cap V^{\prime} \neq \emptyset$. Hence, for any policy $\mu$, the probability of satisfying the constraints is maximized when $S \cap V^{\prime} \neq \emptyset$ for all MECs. 
\end{IEEEproof}
Note that the converse of the lemma may not be true. There may exist policies that maximize the probability of satisfaction and yet do not reach an AMEC $(V^{\prime},\mathcal{A}^{\prime}, P^{\prime})$ with positive probability.


Lemma \ref{lemma:max_reach} implies that in order to formulate the problem of minimizing the ACPC such that the probability of achieving the specifications is maximized, it suffices to ensure that there is at least one state in each AMEC that belongs to $S$. We will show that this is equivalent to a matroid constraint on $S$. Define a partition matroid by $\mathcal{N}_{1} = (V, \mathcal{I})$ where $$\mathcal{I} = \{S : |S \cap V^{\prime}| \leq 1 \ \forall \mbox{ AMEC } \mathcal{M}_{m}^{\prime}, m=1,\ldots,N\}.$$ Let $\mathcal{N}_{k-N}$ denote the uniform matroid with cardinality $(k-N)$. Finally, let $\mathcal{N} = \mathcal{N}_{1} \vee \mathcal{N}_{k-N}$. The following theorem gives the equivalent formulation.

\begin{theorem}
\label{theorem:joint-matroid}
Let $q(S)$ denote the ACPC for set $S$. Then the problem of selecting a set of up to $k$ nodes in order to minimize the ACPC while maximizing reachability probability is equivalent to 
\begin{equation}
\label{eq:joint-equivalent}
\begin{array}{ll}
\mbox{minimize} & q(S) \\
\mbox{s.t.} & S \in \mathcal{N}
\end{array}
\end{equation}
\end{theorem}

\begin{IEEEproof}
Since $q(S)$ is strictly decreasing in $S$, the minimum value  is achieved when $|S| = k$. In order to maximize the probability of satisfying the safety and liveness constraints, $S$ must also contain at least one state in each AMEC, implying that $S$ contains a basis of $\mathcal{N}_{1}$. Hence the optimal set $S^{\ast}$ consists of the union of one state from each AMEC (a basis of $\mathcal{N}_{1}$) and $(k-N)$ other nodes (a basis of $\mathcal{N}_{k-r}$), and hence is a basis of $\mathcal{N}$. Conversely, we have that the optimal solution to (\ref{eq:joint-equivalent}) satisfies the constraint $|S| \leq k$ and contains at least one node in each AMEC, and thus is also a feasible solution to the joint reachability and ACPC problem.
\end{IEEEproof}

Hence by Theorem \ref{theorem:ACPC-sufficient} and Theorem \ref{theorem:joint-matroid},  we can formulate the problem of selecting $S$ to minimize the ACPC as 

\begin{equation}
\label{eq:joint-opt-1}
\begin{array}{ll}
\mbox{minimize} & \max{\lambda} \\
S \in \mathcal{N} & \\
\mbox{s.t.} & \lambda + h(i) \leq 1 + \sum_{j=1}^{n}P(i,a,j)h(j) \\
& + \lambda \sum_{j \notin S}{P(i,a,j)} \ \forall i \in V, a \in A_{i}
\end{array}
\end{equation}

If there are multiple AMECs, then each AMEC $(V_{m}^{\prime}, \mathcal{A}_{m}^{\prime}, P_{m}^{\prime})$ will have a distinct value of average cost per cycle $\lambda_{m}$, which will be determined by $S \cap V_{m}^{\prime}$. 
The problem of minimizing the worst-case ACPC is then given by 
\begin{equation}
\label{eq:joint-opt-2}
\begin{array}{ll}
\mbox{minimize} & \max{\{\lambda_{m}(S \cap V_{m}^{\prime}): m=1,\ldots,N\}} \\
S \in \mathcal{M} & \\
\mbox{s.t.} & \lambda_{m} + h(i) \leq 1 + \sum_{j=1}^{n}P(i,a,j)h(j) \\
& + \lambda_{m} \sum_{j \notin S}{P(i,a,j)} \ \forall i \in V_{m}^{\prime}, a \in A_{i}^{\prime}
\end{array}
\end{equation}
This problem is equivalent to
\begin{equation}
\label{eq:joint-opt-3}
\begin{array}{ll}
\mbox{minimize} & \lambda \\
S,\lambda & \\
\mbox{s.t.} & \lambda + h(i) \leq 1 + \sum_{j=1}^{n}{P(i,a,j)h(j)} \\
& + \lambda \sum_{j \notin S}{P(i,a,j)} \ \forall i, a \in A_{i}^{\prime} \\
& |S| \leq k
\end{array}
\end{equation}
By Proposition \ref{prop:ACPC-submod}, Eq. (\ref{eq:joint-opt-3}) can be rewritten as 
\begin{equation}
\label{eq:joint-opt-4}
\begin{array}{ll}
\mbox{minimize} & \lambda \\
S, \lambda & \\
\mbox{s.t.} & \sum_{m=1}^{N}{r_{\lambda,m}(S \cap V_{m}^{\prime})} = 0 \\
 & |S| \leq k
\end{array}
\end{equation}
where $r_{\lambda,m}(S)$ is the volume of the polytope  $\mathcal{P}(\lambda,S \cap V_{m}^{\prime})$ defined as in Section \ref{subsec:ACPC}  and restricted to the MDP $\mathcal{M}_{m}$. A bijection-based approach, analogous to Algorithm \ref{algo:ACPC}, suffices to approximately solve (\ref{eq:joint-opt-4}). This approach is given as Algorithm \ref{algo:joint-ACPC}.


    \begin{center}
\begin{algorithm}[!htp]
	\caption{Algorithm for selecting a set of states $S$ to jointly optimize ACPC and reachability.}
	\label{algo:joint-ACPC}
	\begin{algorithmic}[1]
		\Procedure{Min\_ACPC\_Max\_Reach}{$V$, $A$, $P$, $R$, $k$}
            \State \textbf{Input}: Set of states $V$, set of actions $A_{i}$ at each state $i$, probability distribution $P$, set of unsafe states $R$,  number of states $k$.
            \State \textbf{Output}: Set of nodes $S$
            \State $\lambda_{max} \leftarrow $ max ACPC for any $v \in \{1,\ldots,n\}$
            \State $\lambda_{min} \leftarrow 0$
            \While{$|\lambda_{max}-\lambda_{min}| > \delta$}
            \State $\lambda \leftarrow \frac{\lambda_{max}+\lambda_{min}}{2}$
            \For{$m=1,\ldots,M$}
            \State $S \leftarrow \emptyset$
                        \While{$r_{\lambda,m}(S) > 0$}
            \State $v^{\ast} \leftarrow \arg\min{\left\{r_{\lambda,m}(S \cup \{v\}): \right. }$ \\
            $\left. v \in \{1,\ldots,n\}\right\}$
            \State $S \leftarrow S \cup \{v^{\ast}\}$
            \EndWhile
            \EndFor
            \If{$|S| \leq k$}
            \State $\lambda_{max} \leftarrow \lambda$
            \Else
            \State $\lambda_{min} \leftarrow \lambda$
            \EndIf
            \EndWhile
            \State \Return{$S$}
		\EndProcedure
	 \end{algorithmic}
\end{algorithm}
\end{center}

The following proposition describes the optimality bounds of Algorithm \ref{algo:joint-ACPC}. 

\begin{proposition}
\label{prop:joint-ACPC-optimality}
Algorithm \ref{algo:joint-ACPC} returns a value of $\lambda$, denoted $\hat{\lambda}$, that satisfies $\hat{\lambda} < \lambda^{\ast}$, where $\lambda^{\ast}$ is the minimum ACPC that can be achieved by any set $S$ satisfying 
\begin{equation}
\label{eq:joint-opt-bound}
|S| \leq k \left(1 + \log{\frac{\sum_{m=1}^{N}{r_{\lambda,m}(\emptyset)}}{\min_{v}{\sum_{m=1}^{N}{r_{\lambda,m}(\hat{S} \setminus \{v\})}}}}\right).
\end{equation} 
\end{proposition}

\begin{IEEEproof}
The proof is analogous to the proof of Theorem \ref{theorem:ACPC-complexity}. The submodularity of  $r_{\lambda,m}(S)$ implies that the set $|\hat{S}|$ is within the bound (\ref{eq:joint-opt-bound})  of the minimum-size set $|S^{\ast}|$ with  $\sum_{m=1}^{N}{r_{\lambda,m}(S)} = 0$.
\end{IEEEproof}

We now turn to the second joint optimization problem, namely, maximizing the reachability probability subject to a constraint $\lambda$ on the average cost per cycle. We develop a two-stage approach. In the first stage, we select a set of input nodes for each AMEC $V_{1},\ldots,V_{N}$ in order to guarantee that the ACPC is less than $\lambda$. In the second stage, we select the set of AMECs to include in order to satisfy the ACPC constraint while minimizing the number of inputs needed.

In the first stage, for each AMEC we approximate the problem 
\begin{equation}
\label{eq:joint-first-stage}
\begin{array}{ll}
\mbox{minimize} & |S_{m}| \\
\mbox{s.t.} & r_{\lambda,m}(S_{m}) = 0
\end{array}
\end{equation}
 We let $c_{m} = |S_{m}|$ denote the number of states required for each AMEC $\mathcal{M}_{m}$.

The second stage problem can be mapped to a maximum reachability problem by defining an MDP $\tilde{M} = (\tilde{V}, \tilde{A}, \tilde{P})$, defined as follows. Let $\mathcal{M}_{1} = (V_{1}^{\prime}, A_{1}^{\prime}, P_{1}^{\prime}), \ldots, \mathcal{M}_{N}=(V_{N}^{\prime},A_{N}^{\prime},P_{N}^{\prime})$ denote the set of AMECs of $\mathcal{M}$ satisfying $V_{i}^{\prime} \cap S \neq \emptyset$.  The node set of the augmented graph is equal to $V \setminus \left(\bigcup_{m=1}^{N}{V_{m}^{\prime}}\right) \cup \{l_{1},\ldots,l_{N}\}$. Here, each $l_{m}$ represents the AMEC $\mathcal{M}_{m}$, so that reaching $l_{m}$ is equivalent to reaching an AMEC $\mathcal{M}_{m}$. The actions for states in $V$ are unchanged, while the states $l_{1},\ldots,l_{N}$ have empty action sets. The transition probabilities are given by 
\begin{displaymath}
\tilde{P}(i,a,j) = \left\{
\begin{array}{ll}
P(i,a,j), & i,j \notin \bigcup_{m=1}^{N}{V_{m}^{\prime}}, a \in A_{i} \\
\sum_{j \in V_{l}^{\prime}}{P(i,a,j)}, & i \notin \bigcup_{s=1}^{N}{V_{s}^{\prime}}, \\
 & j = l_{m}, a \in A_{i} \\
 1, & i=j=l_{m} \\
 0, & i = l_{m}, i \neq j
\end{array}
\right.
\end{displaymath}
In this MDP, the probability of reaching the set $\{l_{1},\ldots,l_{N}\}$ is equal to the probability of satisfying the safety and liveness constraints, and hence maximizing the probability of satisfying the constraints is equivalent to a reachability problem.

 The problem of selecting a subset of states to maximize reachability while satisfying this constraint on $\lambda$ can then be formulated as
\begin{equation}
\label{eq:joint-reachability-max}
\begin{array}{lll}
\mbox{minimize} & \max & \left(\sum_{i=1}^{|\tilde{V}|}{x_{i}} - \lambda\sum_{m \in S}{x_{l_m}}\right)\\
S \subseteq \{1,\ldots,N\},  & \mathbf{x} \in \Pi & \\
\sum_{m \in S}{c_{m}} \leq k & & 
\end{array}
\end{equation}
by analogy to (\ref{eq:reachability-tractable}), where $\Pi$ is defined for the MDP $\tilde{\mathcal{M}}$. The inner optimization problem of (\ref{eq:joint-reachability-max}) is a knapsack problem, and hence is NP-hard and must be approximated at each iteration. In order to reduce the complexity of the problem, we introduce the following alternative formulation. We let $\mathcal{P}_{\lambda}$ denote the polytope satisfying the inequalities 
\begin{eqnarray*}
\mathbf{1}^{T}\mathbf{z} &=& \beta \\
z_{i}(1-\lambda\mathbf{1}\{i \in S\}) &\geq& \sum_{j=1}^{n}{z_{j}P(i,a,j)}
\end{eqnarray*}
By Proposition  \ref{prop:ACPC-submod}, the condition that the reachability probability is at most $\lambda$ is equivalent to $\mbox{vol}(\mathcal{P}(\lambda,S_{m}) = 0$. Letting $r_{\lambda}(S)$ denote the volume of $\mathcal{P}_{\lambda}$ when the set of desired states is $S$, the problem is formulated as 

\begin{equation}
\label{eq:alternative-joint-formulation}
\begin{array}{ll}
\mbox{minimize} & \sum_{m \in T}{c_{m}} \\
\mbox{s.t.} & r_{\lambda}(T) = 0
\end{array}
\end{equation}

Problem (\ref{eq:alternative-joint-formulation}) is a submodular knapsack problem with coverage constraints. An algorithm for solving it is as follows. The set $T$ is initialized to $\emptyset$. At each iteration, find the element $m$ that minimizes $$\frac{c_{m}}{r_{\lambda}(T) - r_{\lambda}(T \cup \{m\})},$$ terminating when the condition $r_{\lambda}(T) = 0$ is satisfied. 

Hence, the overall approach is to select a collection of subsets $\{S_{m} : m=1,\ldots,M\}$, representing the minimum-size subsets to satisfy the ACPC constraint on each AMEC $\mathcal{M}_{m}$. We then select a set of AMECs to include in order to satisfy a desired constraint on reachability while minimizing the total number of inputs. The set $S$ is then given by $$S = \bigcup_{m \in T}{S_{m}}.$$  The optimality gap of this approach is described as follows.

\begin{proposition}
\label{prop:joint-optimality-gap}
The set $T$ chosen by the greedy algorithm satisfies  $$\frac{\sum_{m \in T}{c_{m}}}{\sum_{m \in T^{\ast}}{c_{m}}} \leq 1 + \log{\left\{\frac{1}{r_{\lambda}(\hat{T})}\right\}},$$ where $T^{\ast}$ is the optimal solution to (\ref{eq:alternative-joint-formulation}) and $\hat{T}$ is the set obtained by the greedy algorithm prior to convergence.
\end{proposition}

\begin{IEEEproof}
We have that $r_{\lambda}$ is monotone decreasing and supermodular. Hence, by Theorem 1 of \cite{wolsey1982analysis}, the optimality bound holds.  
\end{IEEEproof}

Combining the optimality bounds yields 
\begin{eqnarray*}
\frac{|S|}{|S^{\ast}|} &=& \frac{\sum_{m \in T}{c_{m}}}{\sum_{m \in T^{\ast}}{c_{m}^{\ast}}} \\
&=& \frac{\sum_{m \in T}{c_{m}}}{\sum_{m \in T^{\ast}}{c_{m}}} \frac{\sum_{m \in T^{\ast}}{c_{m}}}{\sum_{m \in T^{\ast}}{c_{m}^{\ast}}} \\
&\leq& \frac{\sum_{m \in T}{c_{m}}}{\sum_{m \in T^{\ast}}{c_{m}}} \max_{m}{{\left\{\frac{c_{m}}{c_{m}^{\ast}}\right\}}} \\
&\leq& \left(1 + \max_{m}{\log{\left\{\frac{1}{r_{\lambda,m}(\hat{S})}\right\}}}\right)\left(1 + \log{\left\{\frac{1}{r_{\lambda}(\hat{T})}\right\}}\right)
\end{eqnarray*}


\subsection{Optimal Hitting Time}
\label{subsec:optimal-hitting}
In this section, we consider the problem of choosing a set $S$ and policy $\mu$ in order to minimize the hitting time to $S$ in an MDP.
 The hitting time of node $i$ under policy $\mu$, denoted $H(i,\mu,S)$, satisfies  $$H(i,\mu,S) = 1+ \sum_{j=1}^{n}{P(i,\mu(i),j)H(j,\mu,S)}$$ for $i \notin S$ and $H(i,S) = 0$ for $i \in S$. 
 We define $\overline{H}(i,S)$ as the minimum hitting time over all policies $\mu$. Minimizing the hitting time for fixed $S$ is equivalent to solving a stochastic shortest path problem. The solution to this problem is described by the following lemma.
 \begin{lemma}[\cite{bertsekas1995dynamic}, Prop 5.2.1]
 The minimum hitting time $\overline{H}(i,S)$ for set $S$ satisfies
 $$\overline{H}(i,S) = 1 + \min_{a \in A_{i}}{\left\{\sum_{j=1}^{n}{P(i,a,j)\overline{H}(j,S)}\right\}}$$
 and is equivalent to the linear program
\begin{equation}
\label{eq:hitting-time-program}
\begin{array}{ll}
\mbox{maximize} & \mathbf{1}^{T}\mathbf{v} \\
\mbox{s.t.} & v_{i} \leq 1 + \sum_{j=1}^{n}{P(i,a,j)v_{j}} \ \forall i, a \in A_{i} \\
 & v_{i} = 0, \ i \in S \\
\end{array}
\end{equation}
\end{lemma}

The following lemma leads to an equivalent formulation to (\ref{eq:hitting-time-program}).

\begin{lemma}
\label{lemma:hitting-time-condition}
For a given graph $G$, there exists $\theta > 0$ such that the conditions 
\begin{eqnarray}
\label{eq:hitting-time-optimal1}
v_{i} &\leq& 1 + \sum_{j=1}^{n}{P(i,a,j)v_{j}} \ \forall i, a \in A_{i} \\
\label{eq:hitting-time-optimal2}
v_{i} &=& 0 \ \forall i \in S 
\end{eqnarray}
 are equivalent to 
 \begin{equation}
 \label{eq:hitting-time-equivalent}
 v_{i} + \theta \mathbf{1}\{i \in S\}v_{i}  \leq 1 + \sum_{j=1}^{n}{P(i,a,j)v_{j}}.
 \end{equation}
 \end{lemma}
\begin{IEEEproof}
If (\ref{eq:hitting-time-equivalent}) holds, then for $\theta$ sufficiently large we must have $v_{i}=0$ for all $i \in S$. The condition then reduces to $$v_{i} \leq 1 + \sum_{j=1}^{n}{P(i,a,j)v_{j}}$$ for all $i \notin S$ and $a \in A_{i}$, which is equivalent to (\ref{eq:hitting-time-optimal1}).
\end{IEEEproof}

From Lemma \ref{lemma:hitting-time-condition}, it follows that in order to ensure that the optimal hitting time for each node is no more than $\zeta$, it suffices to ensure that for each $\mathbf{v}$ satisfying $\mathbf{1}^{T}\mathbf{v} \geq \zeta$, there exists at least one $i$ and $u$ such that $$1 + \sum_{j=1}^{n}{P(i,a,j)v_{j}} < (1 + \theta \mathbf{1}\{i \in S\})v_{i}.$$ We define the  function $\chi_{v}(S)$ as 
\begin{equation}
\label{eq:chi_v}
\chi_{v}(S) = \left\{
\begin{array}{ll}
1, & \mbox{(\ref{eq:hitting-time-optimal1}) and (\ref{eq:hitting-time-optimal2}) hold } \\
0, & \mbox{else}
\end{array}
\right.
\end{equation}

The following lemma relates the function $\chi_{v}(S)$ to the optimality conditions of Lemma \ref{lemma:hitting-time-condition}.

\begin{lemma}
\label{lemma:hitting-time-sufficient}
The optimal hitting time corresponding to set $S$ is bounded above by $\zeta$ if and only if 
\begin{equation}
\label{eq:chi_zeta}
\chi(S,\zeta) \triangleq \int_{\{\mathbf{v}: \mathbf{1}^{T}\mathbf{v} \geq \zeta\}}{\chi_{v}(S) \ dv} = 0.
\end{equation}
 \end{lemma}

\begin{IEEEproof}
Suppose that $\chi(S,\zeta) = 0$. Then for each $v$ satisfying $\mathbf{1}^{T}\mathbf{v} \geq \zeta$, we have that $\chi_{v}(S,\zeta) = 0$ and hence $$\sum_{j=1}^{n}{P(i,a,j)v_{j}} < (1 + \theta \mathbf{1}\{i \in S\})v_{i}$$ holds for some $i$ and $a \in A_{i}$, and hence there is no $\mathbf{v}$ satisfying the conditions of (\ref{eq:hitting-time-program}) with $\mathbf{1}^{T}\mathbf{v} = \zeta$ for the given value of $S$.
\end{IEEEproof}

The following result then leads to a submodular optimization approach to computing the set $S$ that minimizes the hitting time.

\begin{proposition}
\label{chi_v_supermodular}
The function  $\chi(S,\zeta)$ is supermodular as a function of $S$ for any $\zeta \geq 0$.
\end{proposition}

\begin{IEEEproof}
We first show that $\chi_{v}(S)$ (Eq. (\ref{eq:chi_v})) is supermodular. We have that $\chi_{v}(S) = 1$ if and only if there exists $i \in S$ with $v_{i} \neq 0$, or equivalently, $\mbox{supp}(v) \cap S \neq \emptyset$. If $\chi_{v}(S) = \chi_{v}(S \cup \{u\}) = 0$, then the condition $\mbox{supp}(v) \cap S \neq \emptyset$ already holds, and hence $\chi_{v}(T) = \chi_{v}(T \cup \{u\})$. Since $\chi(S)$ is an integral of supermodular functions (\ref{eq:chi_zeta}), it is supermodular as well. 
\end{IEEEproof}

Furthermore, $\chi(S)$ can be approximated in polynomial time via a sampling-based algorithm \cite{lovasz1993random}. Hence the problem of selecting a set of states $S$ in order to minimize the optimal hitting time can be stated as 

\begin{equation}
\label{eq:opt-hitting-time-submodular}
\begin{array}{ll}
\mbox{minimize} & \zeta \\
\mbox{s.t.} & \min{\{|S| : \chi(S,\zeta) = 0\}} \leq k
\end{array}
\end{equation}

Problem (\ref{eq:opt-hitting-time-submodular}) can be approximately solved using an algorithm analogous to Algorithm \ref{algo:ACPC}, with $r_{\lambda}(S)$ replaced by $\chi(S,\zeta)$ in Lines 9 and 10, and $\lambda$ replaced by $\zeta$ throughout. The following proposition gives optimality bounds for the revised algorithm.

\begin{proposition}
\label{prop:hitting-time-optimality}
The modified Algorithm \ref{algo:ACPC} guarantees that the hitting time satisfies $\overline{H}(i,S) \leq \overline{H}(i,\hat{S})$, where $\hat{S}$ is the solution to $\min{\{\overline{H}(i,S) : |S| \leq \hat{k}\}}$ and $\hat{k}$ satisfies
\begin{equation}
\label{eq:hitting-time-bound}
k = \hat{k}\left(1 + \log{\left\{\frac{\chi(\emptyset)}{\min_{v}{\chi(S \setminus \{v\})}}\right\}}\right).
\end{equation}
\end{proposition}

\begin{IEEEproof}
Since the function $\chi(S,\zeta)$ is supermodular, the number of states $S$ selected by the greedy algorithm to satisfy $\chi(S,\zeta) = 0$ satisfies (\ref{eq:hitting-time-bound}). Hence, for the set $\hat{S}$, since $|S| \leq \hat{k}$, we have that $\overline{H}(i,S) \leq H(i,\hat{S})$.
\end{IEEEproof}

\section{Submodularity of Walk Times with Fixed Distribution}
\label{sec:fixed}

This section demonstrates submodularity of random walk times for walks with a probability distribution that does not depend on the set of nodes $S$. We first state a general result on submodularity of stopping times, and then prove submodularity of hitting, commute, cover, and coupling times as special cases.

\subsection{General Result}
\label{subsec:fixed-general}
Consider a set of stopping times $Z_{1},\ldots,Z_{N}$ for a random process $X_{k}$. Let $W = \{1,\ldots,N\}$, and define two functions $f,g: 2^{W} \rightarrow \mathbb{R}$ by 
\begin{eqnarray*}
f(S) &=& \mathbf{E}\left\{\max{\{Z_{i} : i \in S\}}\right\} \\
g(S) &=& \mathbf{E}\left\{\min{\{Z_{i} : i \in S\}}\right\}
\end{eqnarray*}

We have the following general result.

\begin{proposition}
\label{prop:stopping-time-fixed}
The functions $f(S)$ and $g(S)$ are nondecreasing submodular and nonincreasing supermodular, respectively.
\end{proposition}

\begin{IEEEproof}
For any $S$ and $T$ with $S \subseteq T$, we have that 
\begin{eqnarray*}
 f(T) &=& \mathbf{E}\left\{\max{\left\{\max{\{Z_{i} : i \in S\}},\right.}\right. \\
 && \quad \left. \left. \max{\{Z_{i} : i \in T \setminus S\}}\right\}\right\} \\
 &\geq& \mathbf{E}\left\{\max{\{Z_{i} : i \in S\}}\right\} = f(S),
 \end{eqnarray*}
implying that $f(S)$ is nondecreasing. The proof of monotonicity for $g(S)$ is similar.

To show submodularity of $f(S)$, consider any sample path $\omega$ of the random process $X_{k}$. For any sample path, $Z_{i}$ is a deterministic nonnegative integer. Consider any sets $S,T \subseteq \{1,\ldots,N\}$, and suppose without loss of generality that $\max{\{Z_{i}(\omega) : i \in S\}} \geq \max{\{Z_{i}(\omega) : i \in T\}}$. Hence for any sets $S,T \subseteq \{1,\ldots,N\}$, we have 
\begin{eqnarray*}
\max_{i \in S}{Z_{i}(\omega)} + \max_{i \in T}{Z_{i}(\omega)} &=& \max_{i \in S \cup T}{Z_{i}(\omega)} + \max_{i \in T}{Z_{i}(\omega)} \\
&\geq& \max_{i \in S \cup T}{Z_{i}(\omega)} + \max_{i \in S \cap T}{Z_{i}(\omega)}. 
\end{eqnarray*}
due to submodularity of the max function. 
Now, let $Q = (i_{1},\ldots,i_{N})$, where each $i_{j} \in \{1,\ldots,N\}$, be a random variable  defined by $Z_{i_{1}} \leq Z_{i_{2}} \leq \cdots \leq Z_{i_{N}}$ for each sample path, so that $Q$ is the order in which the stopping times $Z_{i}$ are satisfied. Define $\alpha_{jQ} = \mathbf{E}(Z_{i_{j}}|Q)$.  For any ordering $(i_{1},\ldots,i_{N})$, we have 
\begin{IEEEeqnarray*}{rCl}
\IEEEeqnarraymulticol{3}{l}{
\mathbf{E}\left\{\max_{i \in S}{\{Z_{i}\}} | Q=(i_{1},\ldots,i_{N})\right\}} \\
 &=& \mathbf{E}\left\{\max_{j : i_{j} \in S}{\{Z_{i_{j}}\}} | Q = (i_{1},\ldots,i_{N})\right\} \\
&=& \max_{j : i_{j} \in S}{\mathbf{E}(Z_{i_{j}}|Q)} = \max_{j: i_{j} \in S}{\alpha_{jQ}}.
\end{IEEEeqnarray*}
which is submodular by the same analysis as above. Taking expectation over all the realizations of $Q$, we have 
\begin{eqnarray*}
f(S) &=& \sum_{(i_{1},\ldots,i_{N})}{\mathbf{E}\left\{\max_{i \in S}{\{Z_{i}\}}|Q=(i_{1},\ldots,i_{N})\right.} \\
&& \cdot \left.Pr(Q=i_{1},\ldots,i_{N})\right\} \\
&=& \sum_{(i_{1},\ldots,i_{N})}{\left[\left(\max_{j : i_{j} \in S}{\alpha_{jQ}}\right)Pr(Q=i_{1},\ldots,i_{N})\right]}
\end{eqnarray*}
which is a finite nonnegative weighted sum of submodular functions and hence is submodular. The proof of supermodularity of $g(S)$ is similar. 
\end{IEEEproof}

Proposition \ref{prop:stopping-time-fixed} is a general result that holds for any random process, including random processes that are non-stationary. 
In the following sections, we apply these results and derive tighter conditions for hitting, commute, and cover times.

\subsection{Submodularity  of Hitting and Commute Times}
\label{subsec:hitting}

In this section, we consider the problem of selecting a subset of nodes $S$ in order to minimize the  hitting time to $S$, $H(\pi,S)$, as well as selecting a subset of nodes $S$ in order to minimize the commute time $K(\pi,S)$. The following result is a corollary of Proposition \ref{prop:stopping-time-fixed}.

\begin{lemma}
\label{lemma:hitting-time-submodular}
$H(\pi,S)$ is supermodular as a function of $S$.
\end{lemma}

\begin{IEEEproof}
Let $Z_{i}$ denote the stopping time corresponding to the event $\{X_{k} = i\}$, where $X_{k}$ is a random walk on the graph. Then $H(\pi,S)$ is supermodular by Proposition \ref{prop:stopping-time-fixed}. 
\end{IEEEproof}

The submodularity of $H(\pi,S)$ implies that the following greedy algorithm can be used to approximate the solution to $\min{\{H(\pi,S) : |S| \leq k\}}$. In the greedy algorithm, at each iteration the node $v$ that minimizes $H(\pi, S \cup \{v\})$ is added to $S$ at each iteration. Letting $S^{\ast}$ denote the minimizer of $\{H(\pi, S) : |S| \leq k\}$. Then the set $\hat{S}$ obtained by the greedy algorithm satisfies $$H(\pi, \hat{S}) \leq \left(1-\frac{1}{e}\right)H(\pi,S^{\ast}) + \frac{1}{e}\max_{v}{H(\pi,\{v\})}.$$ An analogous lemma for the commute time is as follows.


\begin{lemma}
\label{lemma:commute-time-supermodular}
For any distribution $u$, the function $K(u,S)$ is supermodular as a function of $S$.
\end{lemma}

\begin{IEEEproof}
Let $Z_{i}$ denote the stopping time corresponding to the event $\{X_{k} = u, X_{l} = i \mbox{ for some } l < k\}$. Then $K(S,u) = \mathbf{E}(\min_{i \in S}{Z_{i}})$, and hence $K(u,S)$ is supermodular as a function of $S$.
\end{IEEEproof}

Lemma \ref{lemma:commute-time-supermodular} can be extended to distributions $\pi$ over the initial state $u$ as $K(\pi,S) = \sum_{u}{K(u,S)\pi(u)}$. The function $K(\pi,S)$ is then a nonnegative weighted sum of supermodular functions, and hence is supermodular.

\subsection{Submodularity of Cover Time}
\label{subsec:cover}

The submodularity of the cover time is shown as follows.

\begin{figure*}
\centering
$\begin{array}{ccc}
\includegraphics[width=2.25in]{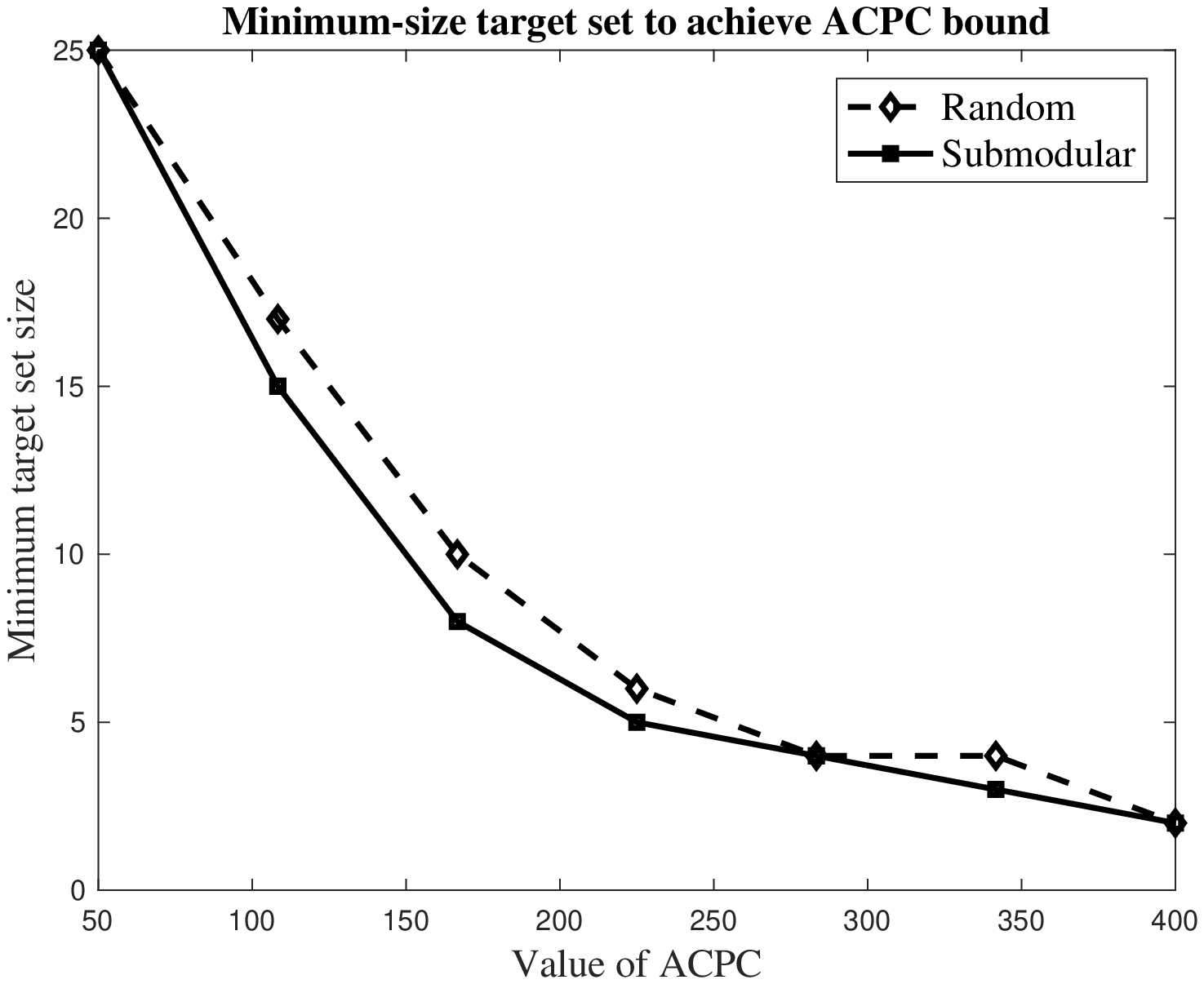} &
\includegraphics[width=2.25in]{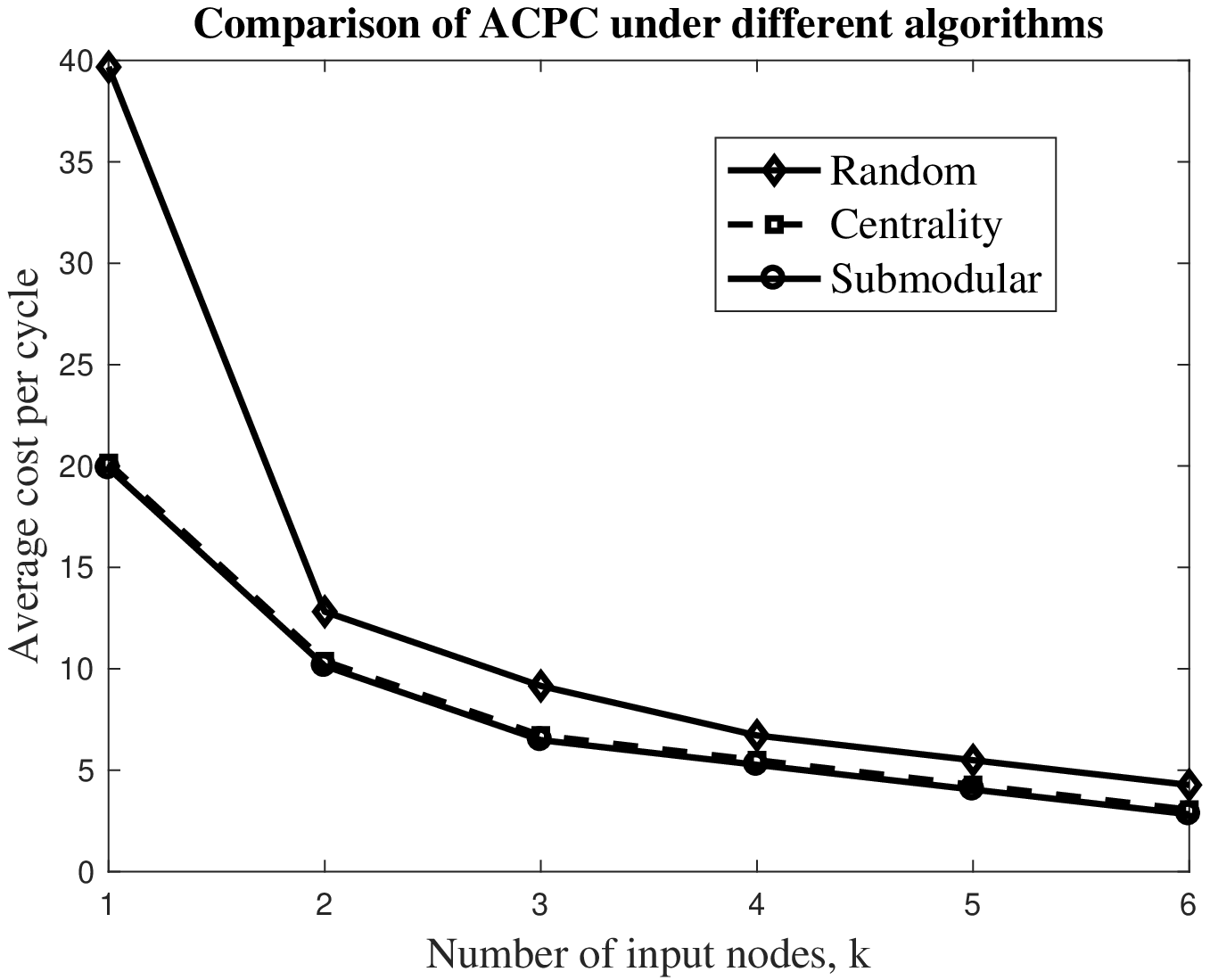} &
\includegraphics[width=2.25in]{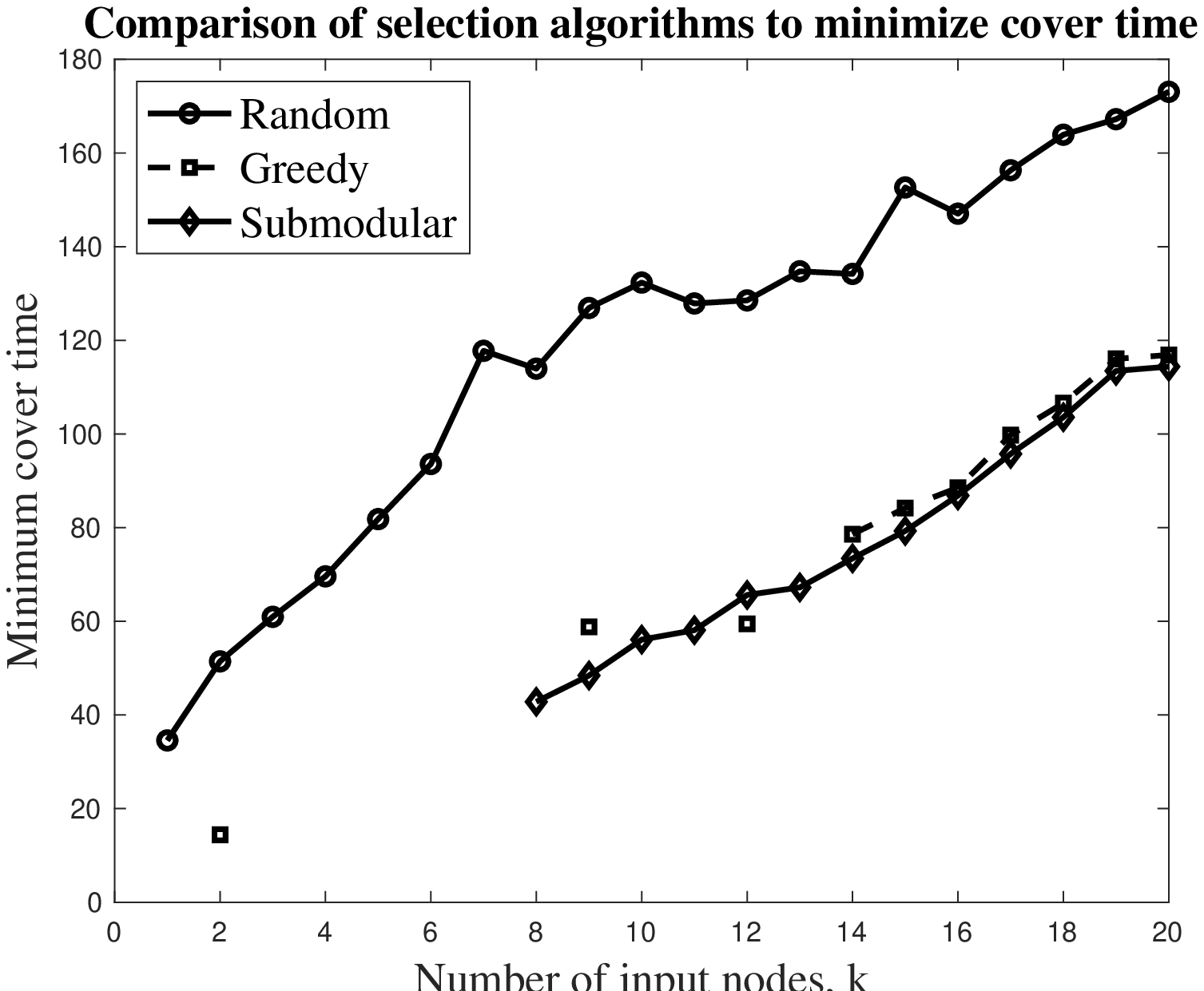} \\
(a) & (b) & (c)
\end{array}$
\caption{Numerical evaluation of submodular optimization of random walk times. (a) Minimum number of input nodes for minimizing average cost per cycle of a uniformly random MDP. (b) Performance of Algorithm \ref{algo:ACPC} for selecting a given number of input nodes to minimize the average cost per cycle. (c) Comparison of minimum cover time using random, greedy, and submodular optimization algorithms.}
\label{fig:fixed}
\end{figure*}

\begin{proposition}
\label{prop:cover-time}
The cover time $C(S)$ is nondecreasing and submodular as a function of $S$.
\end{proposition}

\begin{IEEEproof}
The result can be proved by using Proposition \ref{prop:stopping-time-fixed} with the set of events $\{Z_{i} : i \in S\}$ where the event $Z_{i}$ is given as $Z_{i} = \{X_{k} = i\}$. 

An alternative proof is as follows. Let $S \subseteq T \subseteq V$, and let $v \in V \setminus T$. The goal is to show that $$C(S \cup \{v\}) - C(S) \geq C(T \cup \{v\}) - C(T).$$ Let $\tau_{v}(S)$ denote the event that the walk reaches $v$ before reaching all nodes in the set $S$, noting that $\tau_{v}(T) \subseteq \tau_{v}(S)$. Let $Z_{S}$, $Z_{T}$, and $Z_{v}$ denote the times when $S$, $T$, and $v$ are reached by the walk, respectively. We prove that the cover time is submodular for each sample path of the walk by considering different cases.

In the first case, the walk reaches  node  $v$ before reaching all nodes in $S$. Then $C(S) = C(S \cup \{v\})$ and $C(T) = C(T \cup \{v\})$, implying that submodularity holds trivially.  In the second case, the walk reaches node $v$ after reaching all nodes in $S$, but before reaching all nodes in $T$. In this case, $C(S \cup \{v\}) - C(S) = Z_{v} - Z_{S}$, while $C(T \cup \{v\})-C(T) = 0$. In the last case, the walk reaches $v$ after reaching all nodes in $T$. In this case, $$C(S \cup \{v\}) - C(S) = Z_{v} - Z_{S} \geq Z_{v} - Z_{T} = C(T \cup \{v\}) - C(T),$$ implying submodularity. Taking the expectation over all sample paths yields the desired result.
\end{IEEEproof}

The submodularity of cover time implies that the problem of maximizing the cover time can be approximated up to a provable optimality bound. Similarly, we can select a set of nodes to minimize the cover time by $$\min{\{C(S) - \psi |S| : S \subseteq V|\}}.$$ The cover time, however, is itself computationally difficult to approximate. Instead, upper bounds on the cover time can be used. We have the following preliminary result.

\begin{proposition}[\cite{levin2009markov}, Prop. 11.4]
\label{prop:Matthews}
For any set of nodes $A$, define $t_{min}^{A} = \min_{a,b \in A, a \neq b}{H(a,b)}$. Then the cover time $C(S)$ is bounded by $$C(S) \geq \max_{A \subseteq S}{\left\{t_{min}^{A}\left(1+ \frac{1}{2} + \cdots + \frac{1}{|A|-1}\right)\right\}}.$$
\end{proposition}


Define $c(k) = 1 + \frac{1}{2} + \cdots + \frac{1}{|A|-1}$, and define $\hat{f}(S)$ by $$\hat{f}(S) = \max_{A \subseteq S}{\left\{c(|A|)\min_{\stackrel{b \in A}{a \in V}}{H(a,b)}\right\}}.$$ The approximation $\hat{f}(S)$ can be minimized as follows. We first have the following preliminary lemma. 

\begin{lemma}
\label{lemma:Matthews-approx}
The function $f^{\prime}(S)$ is equal to $$\hat{f}(S) = \max_{k=1,\ldots,|S|}{\alpha_{k}c(k)},$$ where $\alpha_{k}$ is the $k$-th largest value of $H(a,b)$ among $b \in S$.
\end{lemma}

\begin{IEEEproof}
Any set $A$ with $|A| = k$ will have the same value of $c(A)$. Hence it suffices to find, for each $k$, the value of $A$ that maximizes $t_{min}^{A}$ with $|A| = k$. That maximizer is given by the $k$ elements of $S$ with the largest values of $\min_{a \in V}{H(a,b)}$, and the corresponding value is $\alpha_{k}$.
\end{IEEEproof}

By Lemma \ref{lemma:Matthews-approx}, in order to select the minimizer of $\hat{f}(S) - \lambda|S|$, the following procedure is sufficient. For each $k$, select the $k$ elements of $S$ with the smallest value of $\min{\{H(a,b) : a \in V\}}$, and compute $\beta_{k}c(k) - \lambda k$, where $\beta_{k}$ is the $k$-th smallest value of $\min_{a \in V}{H(a,b)}$ over all $b \in S$. 

In addition, we can formulate the following problem of minimizing the probability that the cover time is above a given threshold. The value of $Pr(C(S) > L)$ can be approximated by taking a set of sample paths $\omega_{1},\ldots,\omega_{N}$ of the walk and ensuring that $C(S ; \omega_{i}) > L$ in each sample path. This problem can be formulated as $$Pr(C(S) > L) \approx \frac{1}{N}\sum_{i=1}^{N}{\mathbf{1}\{C(S ; \omega_{i}) > L\}}.$$ The function $\mathbf{1}\{C(S ; \omega_{i}) > L\}$ is increasing and submodular, since it is equal to $1$ if there is a node in $S$ that is not reached during the first $L$ steps of the walk and $0$ otherwise. Hence the problem of minimizing the probability that the cover time exceeds a given threshold can be formulated as $$\min{\left\{Pr(C(S) > L\ - \psi |S| : S \subseteq V\right\}}$$ and solved in polynomial time.

\section{Numerical Results}
\label{sec:simulation}
We evaluated our approach through numerical study using Matlab. We simulated both the fixed and optimal distribution cases. In the case of fixed distribution, our goal was to determine how the minimum cover time varied as a function of the number of input nodes and the network size. We generated an Erdos-Renyi random graph $G(N,p)$, in which there is an edge $(i,j)$ from node $i$ to node $j$ with probability $p$, independent of all other edges, and the total number of nodes is $N$. The value of $N$ varied from $N=10$ to $N=50$. 
 
 
 In the MDP case, we simulated the average cost per cycle (ACPC) problem. We first considered a randomly generated MDP, in which each state $i$ had four actions and the probability distribution $P(i,a,\cdot)$ was generated uniformly at random. We considered the problem of selecting a minimum-size set of states $S$ in order to satisfy a given bound on ACPC. The results are shown in Figure \ref{fig:fixed}(a). We found that the submodular approach outperformed a random selection heuristic even for the relatively homogeneous randomly generated MDPs.  We also found that the number of states required to achieve a given bound on ACPC satisfied a diminishing returns property consistent with the submodular structure of ACPC.

 We then considered a lattice graph. The set of actions corresponded to moving left, right, up, or down. For each action, the walker was assumed to move in the desired direction with probability $p_{c}$, and to move in a uniformly randomly chosen direction otherwise. If an ``invalid'' action was chosen, such as moving up from the top-most position in the lattice, then a feasible step was chosen uniformly at random.
 
Figure \ref{fig:fixed}(b) shows a comparison between three algorithms. The first algorithm selects a random set of $k$ nodes as inputs. The second algorithm selects input nodes via a centrality-based heuristic, in which the most centrally located nodes are chosen as inputs. The third algorithm is the proposed submodular approach (Algorithm \ref{algo:ACPC}). We found that the submodular approach slightly outperformed the centrality-based method while significantly improving on random selection.

 Figure \ref{fig:fixed}(c) compares the optimal selection algorithm with a greedy heuristic and random selection of inputs. We found that the greedy algorithm closely approximates the optimum at a lower computation cost, while both outperformed the random input selection.

\section{Conclusion}
\label{sec:conclusion}
This paper studied the time required for a random walk to reach one or more nodes in a target set. We demonstrated that the problem of selecting a set of nodes in order to minimize random walk times including commute, cover, and hitting times has an inherent submodular structure that enables development of efficient approximation algorithms, as well as optimal solutions for some special cases as stated below.

We considered two cases, namely, walks with fixed distribution, as well as walks in which the distribution is jointly optimized with the target set by selecting a control policy in order to minimize the walk times. In the first case, we developed a unifying framework for proving submodularity of the walk times, based on proving submodularity of selecting a subset of stopping times, and derived submodularity of commute, cover, and hitting time as special cases. As a consequence, we showed that a set of nodes that minimizes the cover time can be selected using only polynomial number of evaluations of the cover time, and further derived solution algorithms for bounds on the cover time.

In the case where the distribution and target set are jointly optimized, we investigated the problems of maximizing the probability of reaching the target set, minimizing the average cost per cycle of the walk, as well as joint optimization of these metrics. We proved that the former problem admits a relaxation that can be solved in polynomial time, while the latter problem can be approximated via submodular optimization methods. In particular, the average cost per cycle can be minimized by minimizing the volume of an associated linear polytope, which we proved to be a supermodular function of the input set.


\bibliographystyle{IEEEtran}
\bibliography{TAC17}

\begin{IEEEbiography}[{\includegraphics[width=1in,height=1.25in,clip,keepaspectratio]{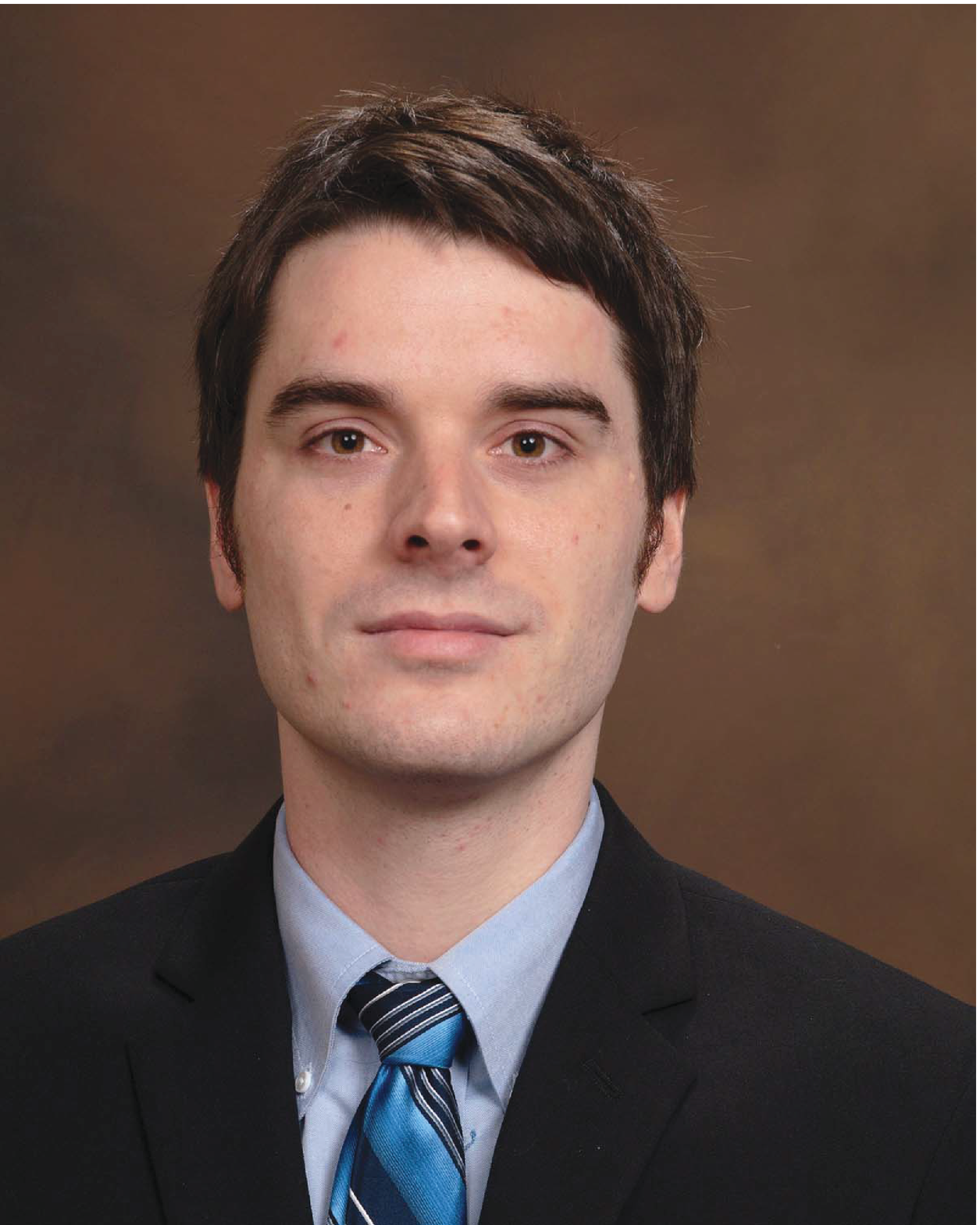}}]{Andrew Clark}(M'15)
is an Assistant Professor in the Department of Electrical and Computer Engineering at Worcester Polytechnic Institute. He received the B.S. degree in Electrical Engineering and the M.S. degree in Mathematics from the University of Michigan - Ann Arbor in 2007 and 2008, respectively. He received the Ph.D. degree from the Network Security Lab (NSL), Department of Electrical Engineering, at the University of Washington - Seattle in 2014. He is author or co-author of the IEEE/IFIP William C. Carter award-winning paper (2010), the WiOpt Best Paper (2012), and the WiOpt Student Best Paper (2014), and was a finalist for the IEEE CDC 2012 Best Student-Paper Award. He received the University of Washington Center for Information Assurance and Cybersecurity (CIAC) Distinguished Research Award (2012) and Distinguished Dissertation Award (2014). He holds a patent in privacy-preserving constant-time identification of RFID. His research interests include control and security of complex networks, submodular optimization, control-theoretic modeling of network security threats, and deception-based network defense mechanisms.
\end{IEEEbiography}

\begin{IEEEbiography}[{\includegraphics[width=1in,height=1.25in,clip,keepaspectratio]{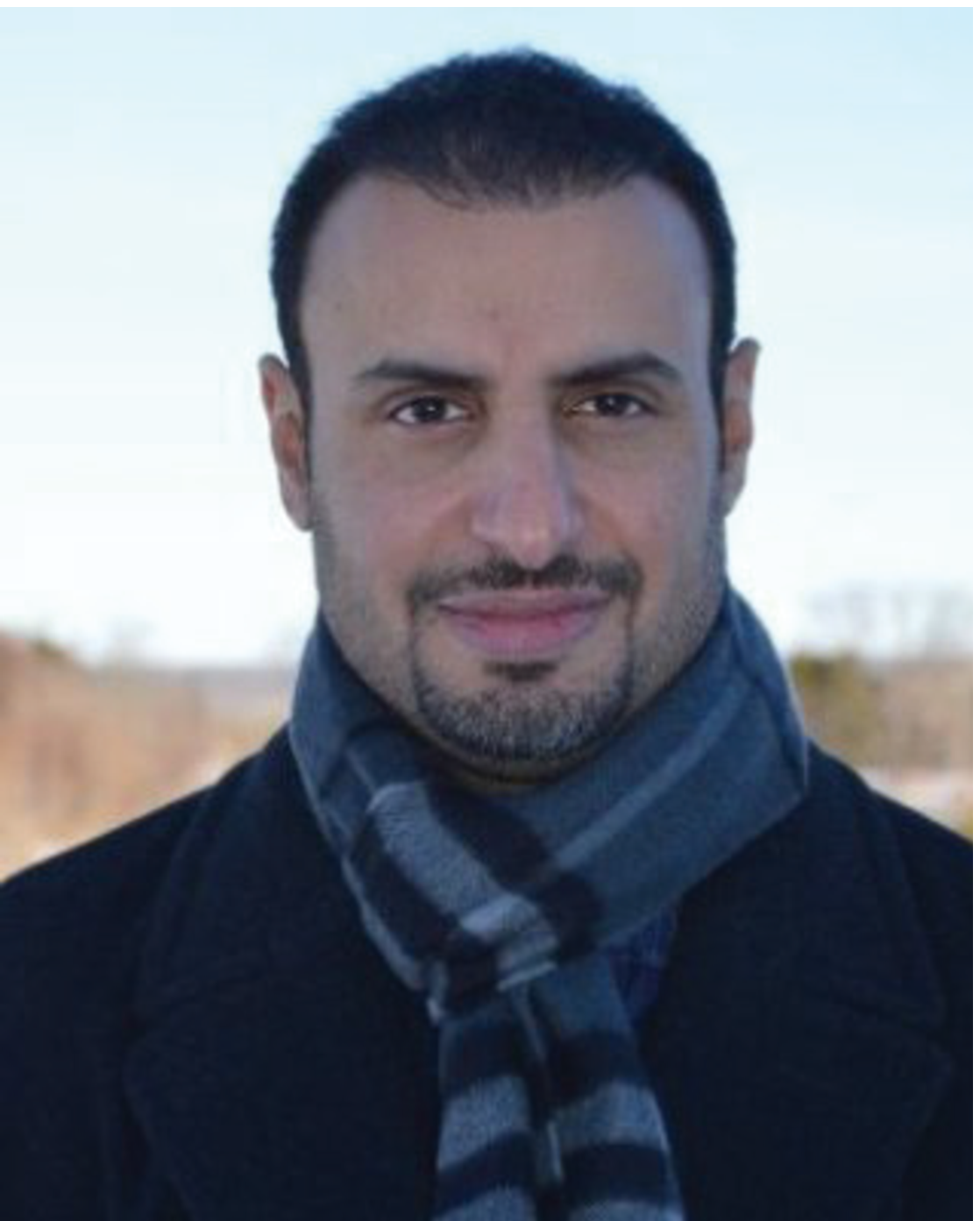}}]{Basel Alomair}(SM'16)
is an Associate Professor and Founding Director of the National Center for Cybersecurity Technology (C4C) in King Abdulaziz City for Science and Technology (KACST), an Affiliate Professor and co-director of the Network Security Lab (NSL) at the University of Washington-Seattle, an Affiliate Professor at King Saud University (KSU), and an Information Security Officer at the Technology Control Company (TCC). He received the B.S. degree in Electrical Engineering from the King Saud University and the M.S. degree in Electrical and Computer Engineering from the University of Wisconsin - Madison in 1998 and 2003, respectively. He received the Ph.D. degree from the Network Security Lab, Department of Electrical Engineering, at the University of Washington - Seattle in 2011. He was recognized by the IEEE Technical Committee on Fault-Tolerant Computing (TC-FTC) and the IFIP Working Group on Dependable Computing and Fault Tolerance (WG 10.4) with the 2010 IEEE/IFIP William Carter Award for his significant contributions in the area of dependable computing. His research in information security was recognized with the 2011 Outstanding Research Award from the University of Washington. He was also the recipient of the 2012 Distinguished Dissertation Award from the University of Washington’s Center for Information Assurance and Cybersecurity (UW CIAC), and he was a co-author of the 2014 WiOpt Best Student Paper Award.
\end{IEEEbiography}

\begin{IEEEbiography}[{\includegraphics[width=1in,height=1.25in,clip,keepaspectratio]{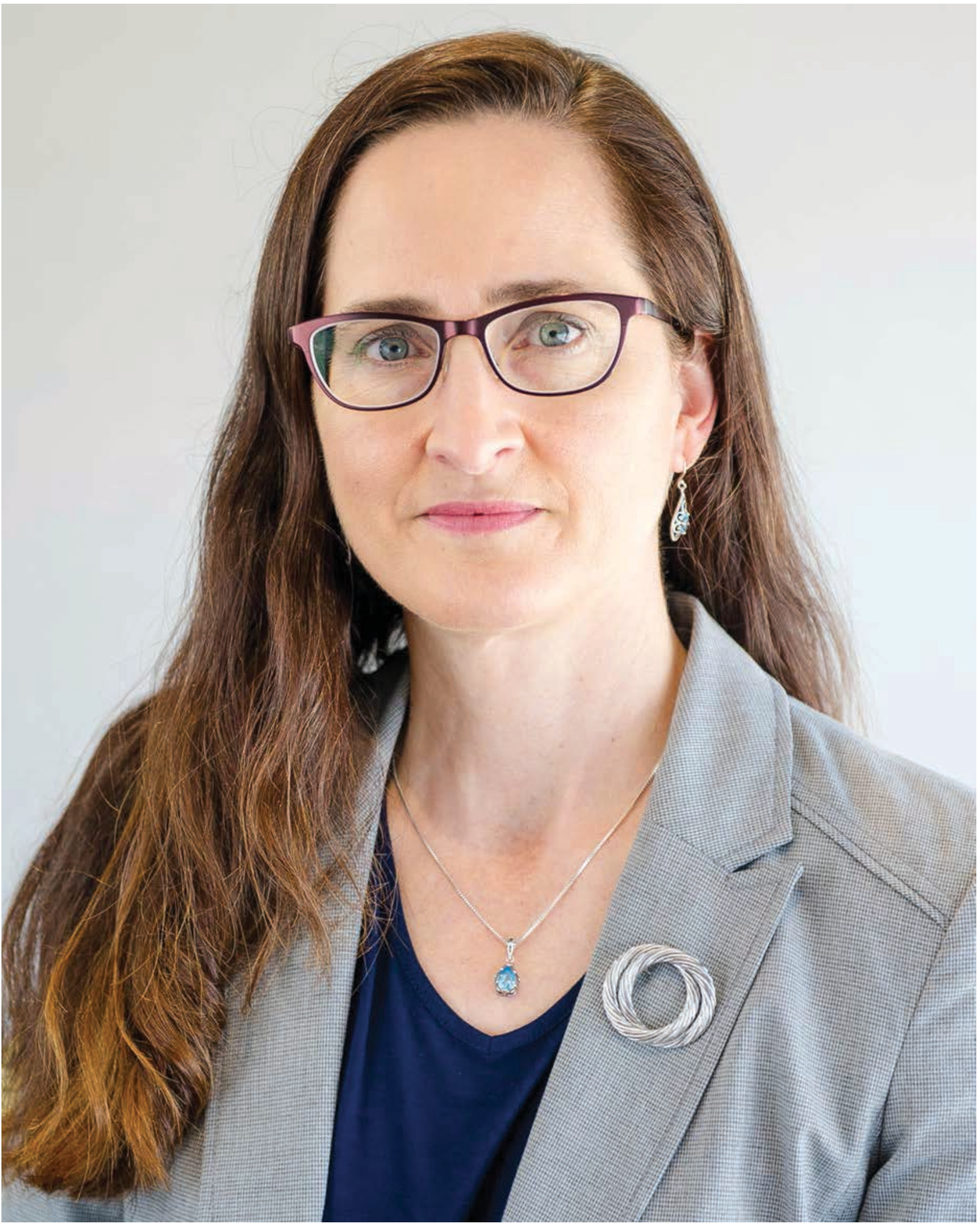}}]{Linda Bushnell}(F'17)
is a Research Professor and the Director of the Networked Control Systems Lab in the Electrical and Computer Engineering Department at the University of Washington - Seattle. She received the B.S. degree and M.S. degree in Electrical Engineering from the University of Connecticut - Storrs in 1985 and 1987, respectively. She received the M.A. degree in Mathematics and the Ph.D. degree in Electrical Engineering from the University of California - Berkeley in 1989 and 1994, respectively. Her research interests include networked control systems, control of complex networks, and secure-control. She was elected a Fellow of the IEEE for her contributions to networked control systems. She is a recipient of the US Army Superior Civilian Service Award, NSF ADVANCE Fellowship, and IEEE Control Systems Society Distinguished Member Award. She is currently an Associate-Editor for \emph{Automatica} and \emph{IEEE Transactions on Control of Network Systems} and Series Editor for the Springer series \emph{Advanced Textbooks in Control and Signal Processing.} For the IEEE Control Systems Society (CSS), she is currently a Distinguished Lecturer, Chair of the Women in Control Standing Committee, and Liaison to IEEE Women in Engineering (WIE). She is also the Treasurer of the American Automatic Control  Council (AACC) and a Member of the International Federation of Automatic Control (IFAC) Technical Board.
\end{IEEEbiography}

\begin{IEEEbiography}[{\includegraphics[width=1in,height=1.25in,clip,keepaspectratio]{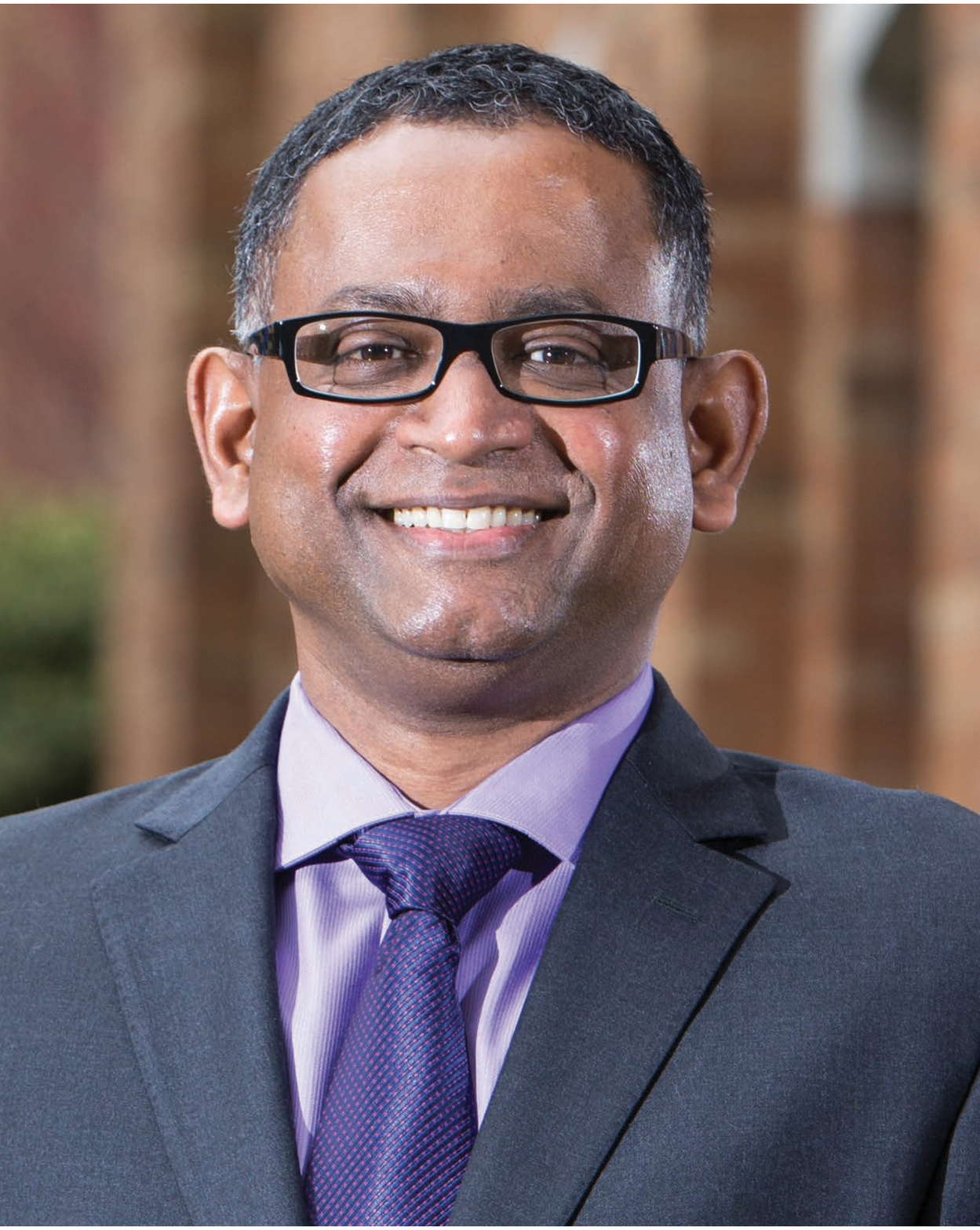}}]{Radha Poovendran}(F'15)
is a Professor and the Chair of the Electrical and Computer Engineering Department at the University of Washington (UW). He is the Director of the Network Security Lab (NSL) at the University of Washington. He is the Associate Director of Research of the UW Center for Excellence in Information Assurance Research and Education. He received the B.S. degree in Electrical Engineering and the M.S. degree in Electrical and Computer Engineering from the Indian Institute of Technology- Bombay and University of Michigan - Ann Arbor in 1988 and 1992, respectively. He received the Ph.D. degree in Electrical and Computer Engineering from the University of Maryland - College Park in 1999. His research interests are in the areas of wireless and sensor network security, control and security of cyber-physical systems, adversarial modeling, smart connected communities, control-security, games-security and information theoretic security in the context of wireless mobile networks. He is a Fellow of the IEEE for his contributions to security in cyber-physical systems. He is a recipient of the NSA LUCITE Rising Star Award (1999), National Science Foundation CAREER (2001), ARO YIP (2002), ONR YIP (2004), and PECASE (2005) for his research contributions to multi-user wireless security. He is also a recipient of the Outstanding Teaching Award and Outstanding Research Advisor Award from UW EE (2002), Graduate Mentor Award from Office of the Chancellor at University of California - San Diego (2006), and the University of Maryland ECE Distinguished Alumni Award (2016). He was co-author of award-winning papers including IEEE/IFIP William C. Carter Award Paper (2010) and WiOpt Best Paper Award (2012). He holds eight patents in wireless and aviation security.
\end{IEEEbiography}

\end{document}